\documentclass[a4paper,11pt]{article}

\usepackage{jheppub}

\usepackage{amsmath, amsthm, amsfonts}
\usepackage[spanish,activeacute,english]{babel}
\usepackage{amsmath}%
\usepackage{amsfonts}%
\usepackage{amssymb}%
\usepackage{graphicx}
\usepackage{cancel}
\usepackage{appendix}
\usepackage{graphicx}
\usepackage{vmargin}
\usepackage{makeidx}
\usepackage{float}
\usepackage{subfig}
\usepackage{braket}
\usepackage{dsfont}
\usepackage{color}
\usepackage{fancybox}
\def\nn{\nonumber}

\theoremstyle{definition}

\theoremstyle{remark}

\title{Non Abelian dual of the resolved conifold gauged linear sigma model}
\author{Nana Geraldine Cabo Bizet$^{a,b}$  \footnote{e-mail address: {\tt nana@fisica.ugto.mx}} , Yulier Jim\'enez Santana  $^{a}$\footnote{e-mail address: {\tt yulier@fisica.ugto.mx}}, Roberto Santos-Silva$^{c}$  \footnote{e-mail address: {\tt roberto.santos@academicos.udg.mx}}\\
[4mm]
{\small \em \it  $^{a}$Departamento de F\'isica, Divisi\'on de Ciencias e Ingenier\'{\i}as, \\ 
Universidad de Guanajuato, Loma del Bosque 103, \\ C.P. 37150, Le\'on, Guanajuato, M\'exico.}\\
[4mm]

{\small \em \it $^{b}$Institut f\"ur Theoretische Physik, ETH Z\"urich Wolfgang-Pauli-Straße 27, 8093,
Z\"urich, Switzerland.}\\
[4mm]

{\small \em \it $^{c}$Departamento Ciencias Naturales y Exactas, CUValles, \\ Universidad de Guadalajara,
Carretera Guadalajara-Ameca Km. 45.5, \\ C.P. 46600, Ameca, Jalisco M\'exico.}\\
[4mm]

}
\date{{\today}}

\abstract{We  consider a $U(1)$ Gauged Linear Sigma Model (GLSM) with $(2,2)$ supersymmetry,
 leading to a susy vacua of the resolved conifold. It possesses the non-Abelian global symmetry $SU(2)\times SU(2)$.
A non-Abelian T-duality can be constructed which can be described by gauging the global non-Abelian symmetry.
This leads to a dual action, in terms of the dual model K\"ahler and superpotential terms,
which include twisted chiral superfields dependence. Comparing the effective potentials
for the $U(1)$ fields on the original and the dual models we determine the instanton corrections to the dual action.
We obtain the supersymmetry vacua solution of the dual model, in three cases: 
first in an Abelian direction inside $SU(2)\times SU(2)$, then for an Abelian direction considering
instanton corrections and finally for a semi-chiral non Abelian vector superfield. The dual geometry for all of these cases (SUSY vacua space) is
given by $T^5\times \mathbb{R}$. However the instanton corrections restrict the Fayet-Iliopoulos term to
be zero and fix the $\theta$ term, which implies that the duality is established for the case of the singular conifold. We also discuss the $U(1)_A$ R-symmetry in the dual model and its possible role.}

\begin{document}
\maketitle

 
\section{Introduction}
Two dimensional Gauge Linear Sigma Models (GLSM) are important tools for the study of  Non Linear Sigma Models (NLSM) arising from string theory  \cite{Witten:1993yc}. A field theory constitutes a sigma model if it describes a set of fields confined on a certain manifold \cite{GellmannLevy}. A non linear sigma model (NLSM) is defined by a map of scalar fields with coordinates on a base manifold and taking values on a target manifold. That is: $X:\Sigma\rightarrow\mathcal{M}$, where $\Sigma$ is the base manifold and $\mathcal{M}$ is the target geometry,  and an integration is performed over $\Sigma$. $X$
is defined as the sets of fields which take values on the base manifold.
Strings propagating in space-time constitute NLSMs. If the target manifold is linear, i.e. the metric
on $\mathcal{M}$ is constant, then the sigma model is a linear sigma model. Gauged linear sigma models (GLSM) are sigma models
when in addition a gauge symmetry exists.
Starting with a supersymmetric 2D GLSM with a given geometry of SUSY vacua under the Renormalization Group (RG) flow, the infrared limit should lead to a string NLSM with a target of this geometry \cite{Hori:2000kt,Hori:2006dk}.  The geometry of the target space in string theory can thus be
studied by analyzing the manifold of supersymmetric vacua $\mathcal{M}_{susy}$ of a given 2D SUSY GLSM.
In particular an important idea we employ is that mirror symmetry between Calabi-Yau manifolds can be understood from GLSMs with Abelian T-duality \cite{Strominger:1996it}.
In the description of Morrison and Plesser \cite{Morrison:1995yh}, mirror symmetry is understood as a T-duality between two GLSMs.   In the description of Hori and Vafa $U(1)^N$ 2D (2,2) GLSMs  and their duals are explored \cite{Hori:2000kt} with the inclusion
of instanton corrections in the dual models, which are connected to vortices on the original model. In these models the string target spaces are Calabi-Yau complete intersection manifolds.
They possess a remnant global symmetry given by powers of $U(1)$s, which can be larger if the chiral fields charges coincide.
The procedure of T-dualization can be interpreted by gauging the global symmetry of a given model and adding Lagrange multipliers \cite{Giveon:1993ai,CaboBizet:2017fzc}. Mirror symmetry continues to be nowadays an actively studied field
by physicist and mathematicians, as the recent works in the subject show\cite{Hausel:2021qkb,Berglund:2021hbo,Collins:2021qqo,Larfors:2021pbb,Alvarez-Consul:2020hbl,Acharya:2020xgn}.
On the other hand GLSMs with non-Abelian gauge group are much less explored than GLSM with Abelian gauge group,
even-so it is estimated that the  number of non-complete intersection CYs is higher than the number of
complete intersection CY varieties \cite{Tonoli2001}. In these models there could be a bigger group of global non-Abelian symmetries, which could lead to interesting dualities or geometrical correspondences.
Dualities arising from the mentioned global symmetries could give rise to interesting relations
and mirror symmetry in these geometries \cite{Gu:2018fpm,Chen:2018wep,Poggi:2018ekr,Gu:2019rty,Gu:2020ana,Gu:2020nub}. Also further
dualities of GLSMs have been explored in the literature \cite{Hori:2011pd,Hori:2006dk,Gerhardus:2015sla,Caldararu:2017usq,Chen:2020iyo}.
Furthermore non-complete intersection Calabi-Yaus and their GLSMs
realizations have been as well actively studied \cite{Caldararu:2007tc,Hosono:2007vf,Jockers:2012zr,Kanazawa:2012xya,Caldararu:2017usq,Honma:2018fgw,Gu:2020oeb,Gu:2020ivl}. Also in recent years there has been a great progress
 in understanding Mirror Symmetry from the perspective of the partition functions,
 in particular employing supersymmetric localization techniques \cite{gomis1,doroud1,benini1,Maxfield:2019czc,doroud2,Hiraga:2020lhk}. 


 In \cite{CaboBizet:2017fzc} a description of non-Abelian T-dualities of 2D (2.2) GLSMs was formulated. 
 As in Mirror Symmetry \cite{Hori:2003ic,Batyrev:2020vnd} a particular case of this T-duality occurs when the dual fields are given by twisted chiral representations \cite{rocek84,Caldeira:2018ynv}.  Also there semi-chiral representations \cite{Bogaerts:1999jc,Gates:2007ve,Gates:1983nr,Rocek:1991ps,Bogaerts:1999jc}
 were found to play an important role; they appear as the representations in the dual fields and as a condition on vector superfields they allow
 to integrate the equations of motion to determine the dual model. Here we want to explore these dualities for a 
 GLSM with a non compact CY as a target. This is the $U(1)$ GLSM with four chiral superfield $\Phi_1$, $\Phi_2$,$\Phi_3$ and $\Phi_4$ with $U(1)$
 charges $(1,1,-1,-1)$ respectively.  The model possesses an $SU(2)\times SU(2)$ global symmetry.
 The supersymmetric vacua of this model is the resolved conifold when the Fayet-Iliopoulus (FI) parameter $r$ is finite. This model it is also denoted $\mathcal{O}(-1)+\mathcal{O}(-1)$ over $\mathbb{CP}^1$, and its
 mirror has been studied \cite{Hori:2000kt}. The duality explored
 here should lead to a different geometrical correspondence.
 Starting from it we obtain the dual model by gauging the 
 global non-Abelian $SU(2)\times SU(2)$ symmetry and introducing a gauge vector superfield. Then
 adding to the Lagrangian the new field strength of vector superfield (vsf) multiplied by a Lagrange multiplier field.  Integrating out  the Lagrange multiplier field one obtains a pure gauge configuration which
 leads to the original model (resolved conifold GLSM).  Integrating out the vsf one goes to the dual model.
 The purpose of this paper is the analysis of the dual non-Abelian model,
 for the resolved conifold GLSM. This will be the first time that these non-Abelian T-dualities in GLSMs are explored for a CY vacua manifold. 
 One of the aims of the present work is to explore symmetries which help reducing the
 landscape of effective string theories, the explorations made here could in the future have
 applications for exploring equivalent stringeffective theories \cite{Shukla:2019dqd}.
 
 In Section 1 we present the tools of $(2,2)$ SUSY in two dimensions.  In Section 2 we
 present the $U(1)$ GLSM which has as a target the resolved conifold.
 In Section 3 we obtain the dual model for an $SU(2)\times SU(2)$ non-Abelian
 duality. In Section 4 we solve the duality for an Abelian direction inside of the non-Abelian 
 gauged group. In Section 5 we perform the duality for this Abelian direction inside $SU(2)\times SU(2)$
 by determining the instanton corrections in the dual theory.
 Finally in Section 6 we solve the non-Abelian duality of a restricted vector
 superfield to be semi-chiral. In Section 7 we present a summary of our results
 and conclusions.





\section{Supersymmetry tools}

In this section we present the fundamental ingredients of $\mathcal{N}=(2,2)$ supersymmetry in 2 dimensions. The  2D supersymmetric theory
can be obtained by dimensional reduction from an $\mathcal{N}=1$ 4D theory \cite{Witten:1993yc}. We define the covariant derivatives and their relations. The different
representations  considered here as: chiral, twisted chiral, vector and semi-chiral will be described. 

The superfields depend on two space-time dimensions (lets chose $x^0$ and $x^3$), such that   the operators can be expressed in terms
of those two coordinates.  The supersymmetric covariant derivatives 
$D_{\alpha}=\frac{\partial}{\partial\theta^{\alpha}}+\sigma^{\mu}_{\alpha\dot{\alpha}}\theta^{\dot{\alpha}}\frac{\partial}{\partial x^\mu}$, $\bar{D}_{\dot{\alpha}}=-\frac{\partial}{\partial\bar{\theta}^{\dot{\alpha}}}-\theta^\alpha\sigma^{\mu}_{\alpha\dot{\alpha}}\frac{\partial}{\partial x^\mu}$
satisfy the Clifford algebra:
\begin{align}
\{D_{\alpha},\bar{D}_{\dot{\alpha}}\}&=-2i\sigma^{\mu}_{\alpha\dot{\alpha}}\partial_\mu,\\
\{D_{\alpha},D_{\beta}\}&=\{\bar{D}_{\dot{\alpha}},\bar{D}_{\dot{\beta}}\}=0,
\end{align}
and can be written as:
\begin{align}
D_{\pm}=\frac{\partial}{\partial\theta^{\pm}}-i\bar{\theta}^{\pm}(\partial_0\mp\partial_3),\hspace{0,1cm}\bar{D}_{\pm}=-\frac{\partial}{\partial\bar{\theta}^{\pm}}+i\theta^{\pm}(\partial_0\mp\partial_3).
\end{align}
The algebra can also be expressed explictly as
\begin{align}
\{D_+,\bar{D}_+\}&=2i(\partial_0-\partial_3),\hspace{0,1cm}\{D_-,\bar{D}_-\}=2i(\partial_0+\partial_3),\\
\{D_+,\bar{D}_-\}&=\{D_-,\bar{D}_+\}=\{D_+,D_-\}=\{\bar{D}_+,\bar{D}_-\}=0.
\end{align}
An unconstrained general  superfield $\Psi$ for $\mathcal{N}=1$ in 4D, which upon dimensional reduction transforms in an unconstrained superfield for $\mathcal{N}=(2,2)$ in 2D, is given
in components  by:
\begin{align}
\Psi&=f+\theta^+\chi_++\theta^-\chi_-+\bar{\theta}^+\bar{\xi}^--\bar{\theta}^-\bar{\xi}^+-2\theta^+\theta^-L+2\bar{\theta}^+\bar{\theta}^-M+\sqrt{2}\theta^+\bar{\theta}^-\Delta+\sqrt{2}\theta^-\bar{\theta}^+\bar{\Delta}+\notag\\
&-\theta^+\bar{\theta}^+(w_0-w_3)-\theta^-\bar{\theta}^-(w_0+w_3)-2\theta^+\theta^-\bar{\theta}^+\bar{\mu}^-+2\theta^+\theta^-\bar{\theta}^-\bar{\mu}^++2\bar{\theta}^+\bar{\theta}^-\theta^+\rho_+\notag\\
&+2\bar{\theta}^+\bar{\theta}^-\theta^-\rho_--4\theta^+\theta^-\bar{\theta}^+\bar{\theta}^-G,\label{uncons}
\end{align}
where $f,\chi_{\alpha},\bar\xi_{\alpha},L,M,\Delta,\bar\Delta,w_{\mu},{\bar\mu}^{\cdot{\alpha}},\rho_{\alpha},G$ are the component fields. They will be employed in this work to construct the Lagrange multiplier fields, which upon integration lead to the original theory and that upon integrating the gauged vector superfield lead to the dual fields.
The chiral and anti-chiral superfields are defined by the following constraint
\begin{equation}
\bar{D}_{\pm} \Phi=0, \quad D_{\pm} \bar{\Phi}=0.
\end{equation}
Thus from the general superfield is easy to construct the chiral and anti-chiral superfields through the next relations:

\begin{eqnarray}
\Phi= \bar{D}_{+} \bar{D}_{-} \Psi,\\
\bar{\Phi}=D_{+} D_{-} \Psi,
\end{eqnarray}
explicitly the chiral superfield is expanded as follows
\begin{align}
\Phi=\bar{D}_{+} \bar{D}_{-} \Psi &=\phi(x)+\sqrt{2}\big(\theta^-\psi^+(x)-\theta^+\psi^-(x)\big)+2\theta^-\theta^+F(x) \label{csf}  \\
&-i\theta^-\bar{\theta}^-\partial_{03}^+\phi(x)
-i\theta^+\bar{\theta}^+\partial_{03}^-\phi(x) \nonumber \\ 
&+\frac{2i}{\sqrt{2}}\theta^+\theta^-\left(\bar{\theta}^-\partial_{03}^+\psi^-(x)+\bar{\theta}^+\partial_{03}^-\psi^+(x)\right) +\theta^+\theta^-\bar{\theta}^+\bar{\theta}^-\partial_{03}^2\phi(x). \nn
\end{align}
The derivatives are given by $\partial^{\pm}_{03}=\partial_0\pm\partial_3$. Notice we have changed the notation with respect to the previous formula. Here $\phi$ is the scalar component, $\psi$ is
a fermion and $F$ is an auxiliary field.
It is important to mention that in the process of dualization we obtain a new set of constraints for the dual fields. These constraints mix covariant derivatives and their conjugated; these fields are called twisted and twisted anti-chiral field. They satisfy the following constraint
\begin{eqnarray}
\bar{D}_+ Y=D_- Y=0,\\
D_+ \bar{Y}= \bar{D}_- \bar{Y}=0,
\end{eqnarray}
in analogy to the (anti-)chiral field this constraint is solved by

\begin{eqnarray}
Y= \bar{D}_{+} D_{-} \Psi,\\
\bar{Y}=D_{+} \bar{D}_{-} \Psi.
\end{eqnarray}
This is we employ the unconstrained superfield to obtain the constrained ones. Let us now introduce a method \cite{bagger} developed in \cite{CaboBizet:2017fzc} for twisted representations, based on employing new coordinates (linear combination of space-time and superspace coordinates) given by:

\begin{eqnarray}
\tilde{X}^\mu=x^\mu + i \theta^+ \sigma^\mu_{+ \dot{+}} \bar\theta^{\dot{+}}+ i \bar\theta^{\dot{-}} \sigma^\mu_{- \dot{-}} \theta^{-}, \\
\bar{\tilde{X}}^{\bar\mu} =x^{\bar\mu} - i \theta^+ \sigma^{\bar\mu}_{+ \dot{+}} \bar\theta^{\dot{+}} - i \bar\theta^{\dot{-}} \sigma^{\bar\mu}_{- \dot{-}} \theta^{-},
\end{eqnarray}
where $\tilde{X}$ and $\bar{\tilde{X}}$ are superspace twisted coordinates which also satisfy the twisted constraints. Now we redefine also the Grassman variables, writing in a compact notation 

\begin{eqnarray}
\tilde\theta^\alpha = \left( \tilde\theta^+ \tilde\theta^- \right)= \left( \theta^+ \bar\theta^- \right),\\
\bar{\tilde\theta}^\alpha= \begin{pmatrix} \bar{\tilde\theta}^+ \\ \bar{\tilde\theta}^- \end{pmatrix} = \begin{pmatrix} \bar{\theta}^+ \\ \theta^- \end{pmatrix}. 
\end{eqnarray}
We rewrite previous expressions into a shorter form

\begin{eqnarray}
\tilde{X}^\mu=x^\mu + i\tilde\theta^\alpha \sigma^\mu_{\alpha \dot{\alpha}} \tilde\theta^{\dot{\alpha}}, \\
\bar{\tilde{X}}^{\bar\mu} =x^{\bar\mu} - i\tilde\theta^\alpha \sigma^{\bar\mu}_{\alpha \dot{\alpha}} \tilde\theta^{\dot{\alpha}}.
\end{eqnarray}
With this we can rewrite the twisted (anti)-chiral superfield $Y$ as follows

\begin{eqnarray}
Y(\tilde X)= y(\tilde X)+ \sqrt{2} \tilde\theta^\alpha \tilde\chi_\alpha(\tilde X) +\tilde\theta^\alpha \tilde\theta_\alpha G(\tilde X), \\
\bar{Y}(\tilde X)= \bar{y}(\tilde X)+ \sqrt{2} \bar{\tilde\theta}^\alpha \bar{\tilde\chi}_\alpha(\tilde X) +\bar{\tilde\theta}^\alpha \bar{\tilde\theta}_\alpha \bar{G}(\tilde X).
\end{eqnarray}
The explicit expansion of the twisted chiral field is given by:

\begin{align}
Y=D_-\bar{D}_+\Psi&=\sqrt{2}\bar{\Delta}-\theta^+(2\bar{\mu}^-+i(\partial_0-\partial_3)\chi_-)+\bar{\theta}^-(2\rho_-+i(\partial_0+\partial_3)\bar{\xi}^-)\\
&-\sqrt{2}i\theta^+\bar{\theta}^+(\partial_0-\partial_3)\bar{\Delta}+\sqrt{2}i\theta^-\bar{\theta}^-(\partial_0+\partial_3)\bar{\Delta}+\theta^+\bar{\theta}^-\times\notag\\
&\times(2iw_{03}-(\partial_0^2-\partial_3^2)f-4G)-\theta^+\theta^-\bar{\theta}^-(2i(\partial_0+\partial_3)\bar{\mu}^--(\partial_0^2-\partial_3^2)\chi_-)\notag\\
&-\theta^+\bar{\theta}^+\bar{\theta}^-(2i(\partial_0-\partial_3)\rho_--(\partial_0^2-\partial_3^2)\bar{\xi}^-)-\theta^+\theta^-\bar{\theta}^+\bar{\theta}^-\sqrt{2}(\partial_0^2-\partial_3^2)\bar{\Delta}, \nn
\end{align}
where the following definition holds: $w_{03}\equiv\partial_0w_3-\partial_3w_0$.

Now we consider semi-chiral superfields, those are defined by the followings constraints

\begin{equation}
\bar{D}_+ n_L =0, \quad \bar{D}_- n_R=0.
\end{equation}
They are the left and right semichiral superfields constriction respectively. These fields can also be constructed
in terms of an unconstrained superfield $\Psi$ as

\begin{equation}
n_L= \bar{D}_+ \Psi, \quad n_R=\bar{D}_- \Psi.
\end{equation}
The other set of semi-chiral superfields fulfill the conditions
 \begin{equation}
D_+ \bar n_L =0, \quad D_- \bar n_R=0,
\end{equation}
these are denoted anti-left and anti-right semi-chiral superfields. They can be obtained
in terms of an unconstrained $\Psi$ superfield through

\begin{equation}
\bar n_L= D_+ \Psi, \quad \bar n_R=D_- \Psi.
\end{equation}
The fields that appear in this work are anti-right semi-chiral superfields. They
are constructed in terms of the non-Abelian and real vector superfields $V=V_a T_a$ ( where $T^a$ are the Lie algebra generators in a given representation) as $n_a=V_a/|V|$. This will be crucial in solving
the equations of motion leading to the dual model in Section \ref{semichiral}.

The field strength of the vector superfields $V$ is defined as a twisted chiral superfield given by
\begin{align}
\Sigma=\frac{1}{2}\bar{D}_+(e^{-V}D_-e^V), \hspace{0,1cm}\bar{\Sigma}=\frac{1}{2}D_+(e^{V}\bar{D}_-e^{-V}).
\end{align}
The vector superfield associated to a group direction $V_a$ has the expansion:
\begin{eqnarray}
V^a&=&\theta^+\bar{\theta}^+(v^a_0-v^a_3)+\theta^-\bar{\theta}^-(v^a_0+v^a_3)-\sqrt{2}\theta^+\bar{\theta}^-\sigma^a-\sqrt{2}\theta^-\bar{\theta}^+\bar{\sigma}^a+2i\bar{\theta}^+\bar{\theta}^-\times\notag\\
&\times&(\theta^+\lambda^{-a}-\theta^-\lambda^{+a})+2i\theta^+\theta^-(\bar{\theta}^-\bar{\lambda}^{+a}-\bar{\theta}^+\bar{\lambda}^{-a})+2\theta^-\theta^+\bar{\theta}^+\bar{\theta}^-D^a.
\end{eqnarray}

Next we list the expansions of the $U(1)$ vector superfield $V_0$ and its corresponding field strength superfield $\Sigma_0$ that will be used along the work:
\begin{eqnarray}
V_0&=&\theta^+\bar{\theta}^+(v_0-v_3)+\theta^-\bar{\theta}^-(v_0+v_3)-\sqrt{2}\theta^+\bar{\theta}^-\sigma-\sqrt{2}\theta^-\bar{\theta}^+\bar{\sigma}+2i\bar{\theta}^+\bar{\theta}^-\times\notag\\
&\times&(\theta^+\lambda^--\theta^-\lambda^+)+2i\theta^+\theta^-(\bar{\theta}^-\bar{\lambda}^+-\bar{\theta}^+\bar{\lambda}^-)+2\theta^-\theta^+\bar{\theta}^+\bar{\theta}^-D.\label{V0}\\
\Sigma_0&=&\frac{1}{\sqrt{2}}\sigma_0+i\theta^+\bar{\lambda}_{0}^-+i\bar{\theta}^-\lambda_{0}^++\theta^+\bar{\theta}^-[D_0-i(v_{03})]- \frac{1}{\sqrt{2}} i\bar{\theta}^-\theta^-\partial^+\bar{\sigma}_0\notag\\
&-& \frac{1}{\sqrt{2}} i\theta^+\bar{\theta}^+\partial^-\bar{\sigma}_0+\bar{\theta}^-\theta^-\theta^+\partial^+\bar{\lambda}_{0}^-+\theta^+\bar{\theta}^+\bar{\theta}^-\partial^-\lambda_{0}^+\notag\\
&-& \frac{1}{\sqrt{2}}\theta^+\bar{\theta}^-\theta^-\bar{\theta}^+(\partial^2_0-\partial_3^2)\bar{\sigma}_0,\label{Sigma0}
\end{eqnarray}
The components fields are: the gauge fields $v_0,v_3$, a complex scalar component $\sigma_0$, the fermionic components  $\lambda_0^+,\lambda_0^-$ and the D-term $D_0$ 
constitutes an auxiliary field. Finally let us write some useful properties of the Grassman variables
employed implicitly in the integrations carried out here:
\begin{eqnarray}
\tilde\theta^2&=&\tilde\theta^\alpha \tilde\theta_\alpha =-2 \theta^+ \bar\theta^-\\
\bar{\tilde\theta}^2 & =&\bar{\tilde\theta}_{\dot\alpha} \bar{\tilde\theta}
^{\dot\alpha}=-2 \bar\theta^+ \theta^-,\\
d \tilde\theta^2 &=&-2 d\theta^+ d\bar\theta^-\\
d\bar{\tilde\theta}^2 &=&-2 d\bar\theta^+ d\theta^-.
\end{eqnarray}

\section{The resolved conifold gauged linear sigma model}

In this section we describe an a $U(1)$ 2D (2,2) GLSM with global symmetry $SU(2)\times SU(2)$.  This is a model with four chiral superfields, such that there are two pairs of them grouped by possessing the same charge under the $U(1)$ local symmetry. We obtain
the supersymmetric vacua space, which is given by the resolved conifold.

We start with a (2,2) 2D GLSM with a $U(1)$ gauge group and four chiral superfields $\Phi_1,\Phi_2,\Phi_3,\Phi_4$. 
The chiral multiplets $\Phi_1,\Phi_2,\Phi_3,\Phi_4$ have $U(1)$ charges: $Q_1=1,Q_2=1,Q_3=-1,Q_4=-1$ respectively. The vectors superfield
of the $U(1)$ symmetry is denoted $V_0$ and its field strength by $\Sigma_0=\frac{1}{2}\bar{D}_+D_-V_0$. Due to this
charge degeneracy the chiral superfields constitute doublets under an $SU(2)\times SU(2)$ global symmetry group.
The two doublets are:  $(\Phi_1,\Phi_2)$ and $(\Phi_3,\Phi_4)$. The model has the following action:

\begin{align}
L_0&=\int d^4\theta\Big[\sum_i^4\bar{\Phi}_ie^{2Q_iV_0}\Phi_i-\frac{1}{2e^2}\bar{\Sigma}_0\Sigma_0\Big]+\frac{1}{2}\left(-\int d^2\tilde{\theta}t\Sigma_0+c.c\right),\label{modelo}
\end{align}.
The coefficient $t=r+i\theta$ has as a real part the Fayet–Iliopoulos(FI) parameter $r$ 
and as imaginary part the $\theta$ parameter, $e$ is the $U(1)$ coupling constant. 
The components of the chiral superfield, the vector superfield and the field strength  are given in formulae (\ref{csf}), (\ref{V0}) and (\ref{Sigma0}) respectively.  The components of $\Phi_i$ with $i=1,2,3,4$,
are the scalar $\phi_i$, the fermion $\psi_i$ and the auxiliary field $F_i$ (\ref{csf}).
The components of the vector superfield $V_0$ are the complex scalar $\sigma_0$, the vector $v_{03}$, the fermions $\lambda_+,\lambda_-$ and the auxiliary field $D_0$ (\ref{V0}). 
Classically this theory possesses vector $U(1)_V$ and an axial $U(1)_A$ symmetries. The $U(1)_A$ could be broken by an anomaly, which is avoided
in cases as the present one when the Calabi-Yau condition $\sum_i Q_i=0$ is fulfilled \cite{Witten:1993yc}. The R-charges of the component fields $(\phi_i, \psi_{i,\pm}, F_i, v_{03}, \sigma_0, \lambda_+,\lambda_-, D_0)$ are $((q_v,0), (q_v -1, \mp 1), (q_v -2, 0), (0, 0), (0, 2), (-1, 1), (1, 1), (0, 0))$, such that the first and second entrances are the vector and axial vector charges respectively \cite{cabosantos19}.

The superfield expansions lead to compute the scalar potential obtained for the Lagrangian of the model (\ref{modelo}), which reads as:

\begin{equation}
U=\sum_{i=1}^4F_i\bar{F}_i+D_0Q_i|\phi_i|^2+2|\sigma_0|^2Q_i^2|\phi_i|^2-\frac{1}{8e^2}D_0^2+rD_0.
\end{equation}
The superfield components can be seen in the superfield expansions (\ref{csf}) and (\ref{Sigma0}).

Eliminating the auxiliary fields $F_i$ and $D_0$ with their equations of motion the potential an be written as:
\begin{equation}
U(\phi,\sigma)=2e^2\left(|\phi_1|^2+|\phi_2|^2-|\phi_3|^2-|\phi_4|^2+r\right)^2+2|\sigma_0|^2\left(|\phi_1|^2+|\phi_2|^2+|\phi_3|^2+|\phi_4|^2\right).
\end{equation}

The supersymmetric vacua space determined by $U=0$, 
for the branch $\sigma_0=0$ is given by:
\begin{align}
|\phi_3|^2+|\phi_4|^2-|\phi_1|^2-|\phi_2|^2=r.
\end{align}
The $U(1)$ action applies on the fields $\phi_i$
as $(\phi_1,\phi_2,\phi_2,\phi_4)\sim (e^{i\lambda}\phi_1,e^{i\lambda}\phi_2,e^{-i\lambda}\phi_3,e^{-i\lambda}\phi_4)$ with parameter $\lambda$. This locus constitutes the resolved conifold. If the Fayet–Iliopoulos parameter $r$ goes to zero,
one has the singular conifold. Due
to its properties this model can also be denoted $\mathcal{O}(-1)+ \mathcal{O}(-1)$ over $\mathbb{CP}^1$. 

The mirror model of (\ref{modelo}) \cite{Hori:2000kt} is obtained via four Abelian dualities, along the directions of four $U(1)s$ leading to the dual twisted superfields $X_1,X_2,X_3,X_4$, with twisted superpotential:
\begin{eqnarray}
\widetilde W_{mirror} =X_1 +X_2 +X_3 +X_4, \, \, X_1X_2 = X_3X_4e^t.
\end{eqnarray}
This can be written as:
\begin{eqnarray}
\widetilde W_{mirror} =X_2 +X_3 +X_4 +e^tX_3X_4/X_2.
\end{eqnarray}
This is the mirror sigma model of (\ref{modelo}). In this work we are going to explore a different set of T-dualities. They are constructed along the $SU(2)\times SU(2)$ symmetry directions.

\section{The $SU(2)\times SU(2)$ dual model}

\hspace{0.5cm}In this section starting from a GLSM which leads to the resolved conifold, we take the global symmetry $SU(2)\times SU(2)$ and make it local, adding a vector superfield and Lagrange multiplier fields. In order to obtain the dual Lagrangian of the original model we integrate with respect to the gauged vector superfield.  The master gauged Lagrangian is given by:

\begin{align}
L&=\int d^4\theta\Big[\bar{\Phi}_i\left(e^{2Q^IV_0+V^I}\right)_{ij}\Phi_j+\bar{\Phi}_p\left(e^{2Q^{II}V{_0}+ V^{II}}\right)_{pq}\Phi_q+\text{Tr}\left(\Psi^I\Sigma^I\right)+\text{Tr}\left(\bar{\Psi}^I\bar{\Sigma}^I\right)\notag\\
&+\text{Tr}\left(\Psi^{II}\Sigma^{II}\right)+\text{Tr}\left(\bar{\Psi}^{II}\bar{\Sigma}^{II}\right)-\frac{1}{2e^2}\bar{\Sigma}_0\Sigma_0\Big]+\frac{1}{2}\left(-\int d^2\tilde{\theta}t\Sigma_0+c.c\right), \label{Lmaster}
\end{align}
where the indices are $i,j=1,2$ and $p,q=3,4$. The fields $\Psi^I$, $\Psi^{II}$ are Lagrangian multiplier unconstrained superfields, $\Sigma^I$ and $\Sigma^{II}$, twisted chiral sueprfields, constitute the field strengths of the vector superfields $V^I$ and $V^{II}$ respectively. Here $I$ and
$II$ denote the first and the second $SU(2)$ factors from
the global symmetry.

Integrating the Lagrange multiplier fields $\Psi^I$ and $\Psi^{II}$ we get 
a pure gauge theory with $\Sigma^I=\Sigma^{II}=0$, thus recovering the original Lagrangian $L_0$ \ref{modelo}.
The dual model is obtained by integrating the superfields $V^I$ and $V^{II}$.
For a non-Abelian local symmetry the field strength is given by \cite{Gates:1983nr}
\begin{equation}
\Sigma=\frac{1}{2}\bar{D}_+(e^{-V}D_-e^V), \bar{\Sigma}=\frac{1}{2}D_+(e^{V}\bar{D}_-e^{-V}).
\end{equation}
 The variation of trace terms and the kinetic term with respect to $V^I$ give:
 \begin{align}
 \delta\text{Tr}(\Psi^I\Sigma^I)&=-\frac{1}{2}\text{Tr}\left[\left(\chi^I\tau^I+\tau^I\chi^I+D_-\tau^I\right)\Delta V^I\right].\\
  \delta\text{Tr}(\bar{\Psi}^I\bar{\Sigma}^I)&=-\frac{1}{2}\text{Tr}\left[\left(\bar{\chi}^I\bar{\tau}^I+\bar{\tau}^I\bar{\chi}^I-\bar {\LARGE D}_-\bar{\tau}^I\right)\Delta V^I\right].\\
  \delta(e^{2Q^IV_0+V^I})&=e^{2Q^IV_0}e^{V^I}\Delta V^I,
 \end{align}
 where the gauge invariant vector superfield variation is given by  $\Delta V^I=e^{-V^I}\delta e^{V^I}$,
 and the following definitions apply: 
 \begin{eqnarray}
 \chi^I&=e^{-V^I}D_-e^{V^I}, & \bar{\chi}^I=-e^{V^I}\bar{D}_-e^{-V^I},  \\
 \tau^I&=\bar{D}_+\Psi^I, & \bar{\tau}^I=D_+\bar{\Psi}^I.\nn 
 \end{eqnarray}
 Thus
 one can write the total variation of the Lagrangian with respect to the vector superfield
 of the first $SU(2)$,  $V^I$ ,  as:

\begin{align}
\delta L&=\bar{\Phi}_i\left(e^{2Q^IV_0+V^I}\right)_{ij}\Delta V^I\Phi_j\\
&-\frac{1}{2}\text{Tr}\Big[\Big(e^{-V^I}D_-e^{V^I}\bar{D}_+\Psi^I+\bar{D}_+\Psi^I e^{-V^I}D_-e^{V^I}+D_-\bar{D}_+\Psi^I\Big)\Delta V^I\Big]\notag \nn\\
&-\frac{1}{2}\text{Tr}\Big[\left(-e^{V^I}\bar{D}_-e^{-V^I}D_+\bar{\Psi}^I-\bar{D}_-D_+\bar{\Psi}^I-D_+\bar{\Psi}^Ie^{V^I}\bar{D}_-e^{-V^I}\right)\Delta V^{I\dagger}\Big]=0 \nn\\ 
&=\bar{\Phi}_i\left(e^{2Q^IV_0+V^I}\right)_{ij}\Phi_j-\frac{1}{2}\text{Tr}\Big[\Big(\chi^I\tau^I+\tau^I\chi^I+D_-\tau^I\Big)\Delta V^I\Big]\notag\\
&-\frac{1}{2}\text{Tr}\Big[\left(\bar{\chi}^I\bar{\tau}^I+\bar{\tau}^I\bar{\chi}^I-\bar{D}_-\bar{\tau}^I \right)\Delta V^{I\dagger}\Big]=0.\nn
\end{align}
The variation of the Lagrangian with respect to the vector superfield $V^{II}$ of the
second $SU(2)$ is an exact copy of the previous one:
\begin{align}
\delta L&=\bar{\Phi}_p\left(e^{2Q^{II}V_0+V^{II}}\right)_{pq}\Delta V^{II}\Phi_q-\frac{1}{2}\text{Tr}\Big[\Big(e^{-V^{II}}D_-e^{V^{II}}\bar{D}_+\Psi^{II}+\bar{D}_+\Psi^{II} e^{-V^{II}}D_-e^{V^{II}}\notag\\
&+D_-\bar{D}_+\Psi^{II}\Big)\Delta V^{II}\Big]-\frac{1}{2}\text{Tr}\Big[\Big(-e^{V^{II}}\bar{D}_-e^{-V^{II}}D_+\bar{\Psi}^{II}-\bar{D}_-D_+\bar{\Psi}^{II}\notag\\
&-D_+\bar{\Psi}^{II}e^{V^{II}}\bar{D}_-e^{-V^{II}}\Big)\Delta V^{^{II}\dagger}\Big]=0\notag\\ 
&=\bar{\Phi}_p\left(e^{2Q^{II}V_0+V^{II}}\right)_{pq}\Phi_q-\frac{1}{2}\text{Tr}\Big[\Big(\chi^{II}\tau^{II}+\tau^{II}\chi^{II}+D_-\tau^{II}\Big)\Delta V^{II}\Big]\notag\\
&-\frac{1}{2}\text{Tr}\Big[\left(\bar{\chi}^{II}\bar{\tau}^{II}+\bar{\tau}^{II}\bar{\chi}^{II}-\bar{D}_-\bar{\tau}^{II} \right)\Delta V^{II\dagger}_2\Big]=0.\label{varL}
\end{align}
We take into account that $\Delta V^{I,II}=(\Delta V^{I,II})^{\dagger}$. Now we will write the equations fo motion deduced from previous variations. Let us define:
\begin{eqnarray}
X_a&=\frac{1}{2}(\chi^I\tau^I+\tau^I\chi^I+D_-\tau^I)_a,\, &\bar X_a=\frac{1}{2}(\bar\chi^I\bar\tau^I+\bar\tau^I\bar\chi^I-\bar D_-\bar\tau^I)_a, \\
Y_a&=\frac{1}{2}(\chi^{II}\tau^{II}+\tau^{II}\chi^{II}+D_-\tau^{II})_a,\, &\bar Y_a=\frac{1}{2}(\bar\chi^{II}\bar\tau^{II}+\bar\tau^{II}\bar\chi^{II}-\bar D_-\bar\tau^{II})_a.
\end{eqnarray}
The sunbindex $a$ denotes the coefficient accompanying the generator $T_a$ in the expansion,
such that $X=X_aT_a$ and $Y=Y_a T_a$.
The equations of motion are the following:
\begin{eqnarray}
\bar \Phi_i  e^{2V_0}(e^{V^I})_{ij} (T_a)_{jk}\Phi_k=X_a+\bar X_a,\,  i,j,k=1,2,\, a=1,2,3. \\
\bar \Phi_p  e^{-2V_0}(e^{V^{II}})_{pq} (T_a)_{(q-2)(r-2)}\Phi_r=Y_a+\bar Y_a,\, p,q,r=3,4,\, a=1,2,3,\nn
 \end{eqnarray}
we have taken into account that the $U(1)$ charges are given by  $Q^I=1, Q^{II}=-1$.

\subsection{Computing the dual model Lagrangian}

The results of (4.7) and (4.8) show that in the case of $SU(2)\times SU(2)$
global symmetry the equations of motion for $X_a$ and $Y_a$ have the same structure.
This will be explicit in our calculations following next.

In terms of the $SU(2)$ group generators $\sigma_a$, the vector superfields $V^I$ and $V^{II}$ can be written as
\begin{eqnarray}
V^I=V^I_a\sigma_a, \, \, V^{II}=V^{II}_a\sigma_a.
\end{eqnarray}
Let us focus in one $SU(2)$ factor, as both of them have equal equations of motion, the model (\ref{Lmaster}) can be considered as two copies of the same Lagrangian. So let us write
$V=V_a\sigma_a$ which will be valid to denote $V^I$ and $V^{II}$. The equations of motion for $V_a$ possesses as a component the  
kinetic term:

\begin{eqnarray}
\bar{\Phi}_i\left(e^{2Q^{I}V_0+V^{I}}\right)_{ij}\Delta V^I\Phi_j&=&e^{2QV_0}\bar{\Phi}_i\left(e^{V}\cdot\sigma_a\right)_{ij}\Delta V_a\Phi_j, \label{eom1}\\
&=&D_-\bar{D}_+\Psi^I_a+D_+\bar{D}_-\bar{\Psi}^I_a=X_a+\bar{X}_a, \nn \\
\bar{\Phi}_i\left(e^{2Q^{II}V_0+V^{II}}\right)_{ij}\Delta V^{II}\Phi_j&=&e^{2QV_0}\bar{\Phi}_i\left(e^{V^{II}}\cdot\sigma_a\right)_{ij}\Delta V^{II}_a\Phi_j \label{eom2}\\
&=&D_-\bar{D}_+\Psi^{II}_a+D_+\bar{D}_-\bar{\Psi^{II}}_a=Y_a+\bar{Y}_a.\nn
\end{eqnarray}
Solving this equation for each generator fo $SU(2)$, $a=1,2,3$ we obtain the kinetic term of the chiral superfield in terms of the twisted chiral superfield as \cite{CaboBizet:2017fzc}:

\begin{eqnarray}
L_{kin}&=&\sqrt{(X_1+\bar{X}_1)^2+(X_2+\bar{X}_2)^2+(X_3+\bar{X}_3)^2} \\
&+& \sqrt{(Y_1+\bar{Y}_1)^2+(Y_2+\bar{Y}_2)^2+(Y_3+\bar{Y}_3)^2}\\
&=&\sqrt{(Z^{I}_1)^2+(Z^{I}_2)^2+(Z^{I}_3)^2}
+\sqrt{(Z^{II}_1)^2+(Z^{II}_2)^2+(Z^{II}_3)^2}, \nn \\
Z^{I}_a&=&X_a+\bar X_a= 2 Re(X_a), \nn \\
Z^{II}_a&=&Y_a+\bar Y_a= 2 Re(X_a). \nn
\end{eqnarray}
We have defined the quantity $Z^{I}_a$
as two times the real part of $X_a$. One can define the quantities $|V^{I}|$, $|V^{II}|$ and $\hat{n}_a$, $\hat{m}_a$ in the following way:
\begin{eqnarray}
&V^{I}=|V^{I}|\hat{n}_a\sigma_a, \, \, &V^{II}=|V^{II}|\hat{n}_a\sigma_a\\
&|V^{I}|=\sqrt{V^{I}_aV^{I}_a}, &|V^{II}|=\sqrt{V^{II}_aV^{II}_a}\\ &\hat{n}_a=\frac{V^{I}_a}{\sqrt{V^{I}_aV^{I}_a}},  &\hat{m}_a=\frac{V^{II}_a}{\sqrt{V^{II}_aV^{II}_a}}
\end{eqnarray}
 
 The  expression for $|V^I|$ in terms of the dual twisted chiral superfields $X_a$ and chiral superfields $\Phi_i,i=1,2$ can be computed by solving (\ref{eom1}) to be
\begin{eqnarray}
|V^I|&=&\text{Ln}\Bigg[\left[(n_1\Phi_1-in_2\Phi_1+\Phi_2-n_3\Phi_2)\bar{\Phi}_1-(\Phi_1+n_3\Phi_1+n_1\Phi_2+in_2\Phi_2)\bar{\Phi}_2\right]Z_1\notag\\
&+&iZ_2\left[(n_1\Phi_1-in_2\Phi_1-\Phi_2+n_3\Phi_2)\bar{\Phi}_1-(1+n_3)\Phi_1\bar{\Phi}_2+(n_1+in_2)\Phi_2\bar{\Phi}_2\right]\times\notag\\
&\times&\frac{1}{2e^{2Q_0V_0}\Phi_1\Phi_2(-2n_3\bar{\Phi}_1\bar{\Phi}_2+n_1(\bar{\Phi}_1^2-\bar{\Phi}_2^2)-in_2(\bar{\Phi}_1^2+\bar{\Phi}_2^2)}\Bigg]\nn\\
&=&\text{Ln}\Bigg[\Phi_1\bar{\Phi}_1\left\{Z_1(n_1-in_2+(1-n_3)g-(1+n_3+(n_1+in_2)g)\bar{g}\right]\notag\\
&+&iZ_2\left[(n_1-in_2+(n_3-1)g-(1+n_3)-(n_1+in_2)g)\bar{g}\right\}\times\notag\\
&\times&\frac{1}{2e^{2Q_0V_0}\Phi_1\bar{\Phi}_1\Phi_2\bar{\Phi}_2(-2n_3+n_1\left(\frac{1}{\bar{g}}-\bar{g}\right)-in_2\left(\frac{1}{\bar{g}}+\bar{g}\right)}\Bigg]\nn\\
|V^I|&=&\text{Ln}\Bigg[Z_1\left[(n_1-in_2+(1-n_3)g-(1+n_3+(n_1+in_2)g)\bar{g}\right]\label{VIexp}\\
&+&iZ_2\left[(n_1-in_2+(n_3-1)g-(1+n_3)-(n_1+in_2)g)\bar{g}\right]\times\notag\\
&\times&\frac{1}{-2n_3+n_1\left(\frac{1}{\bar{g}}-\bar{g}\right)-in_2\left(\frac{1}{\bar{g}}+\bar{g}\right)}\Bigg]-\text{Ln}(2|\bar{\Phi}_2|^2)-2Q_0V_0, \nn
\end{eqnarray}
with the function $g$ given by $g=\frac{\Phi_2}{\Phi_1}$. A total analogous expression can be computed for
$V^{II}$. The previous expression can be summarized as:
\begin{eqnarray}
|V^I|&=&\mathcal{K}(X_a,\bar{X}_a,\hat{n}_a)-\text{Ln}(2|\Phi_2|^2)-2Q^{I}V_0,\\
|V^{II}|&=&\mathcal{K}(Y_a,\bar{Y}_a,\hat{m}_a)-\text{Ln}(2|\Phi_4|^2)-2Q^{II}V_0.
\end{eqnarray}

Using the exponential expansion the following expressions were obtained for the $SU(2)$ symmetry \cite{CaboBizet:2017fzc} to have:

\begin{align}
e^{\pm V^I}&=I \text{cosh}|V^I|\pm\hat{n}_a\sigma_a \text{sinh}|V^I|\\
e^{-V}D_-e^{V^I}&=\left(\hat{n}_aD_-|V^I|+D_-\hat{n}_a\text{sinh}|V^I|\text{cosh}|V^I|-\hat{n}_dD_-\hat{n}_bi,\epsilon_{dba}\text{sinh}^2|V^I|\right)\sigma_a.
\end{align}
The analogous expresssion for the second $SU(2)$
factor can be obtained by exchanging $V^I$
by $V^{II}$ and $n_a$ by $m_a$. We will consider later that a semi-chiral condition is satisfied: $D_-\hat{n}_a=0$ (we already know that $\hat{n}_a=\hat{n}_a^{\dagger}$). The expressions containing  $V$ in (\ref{varL}) are reduced to:
 
\begin{align}
e^{-V^{I}}D_-e^{V^{I}}&=\hat{n}_{a}D_-|V^{I}|\sigma_a,\\
e^{V^{I}}\bar{D}_-e^{-V^{I}}&=-\hat{n}_{a}\bar{D}_-|V^{I}|\sigma_a,\nn \\
e^{-V^{II}}D_-e^{V^{II}}&=\hat{m}_{a}D_-|V^{II}|\sigma_a,\\
e^{V^{II}}\bar{D}_-e^{-V^{II}}&=-\hat{m}_{a}\bar{D}_-|V^{II}|\sigma_a.\nn
\end{align}

The more general duality is obtained by considering a real superfield $\hat{n}_a$ without he semi-chirality condition, but the dual theory would be more complicated to work with. Taking into account the previous  ingredients the dual Lagrangian can be written as
\begin{align}
\tilde{L}&=\int d^4\theta\bigg(\sqrt{({Z_1}^I)^2+({Z_2}^I)^2+({Z_3}^I)^2}-\frac{1}{2}\text{Tr}\left(\bar{D}_+\Psi^{I} e^{-V^I}D_-e^{V^I}+D_+\bar{\Psi}^{I} e^{V^I}\bar{D}_-e^{-V^I}\right)\label{Ldual1}\\
&+\sqrt{({Z_1}^{II})^2+({Z_2}^{II})^2+({Z_3}^{II})^2}-\frac{1}{2}\text{Tr}\left(\bar{D}_+\Psi^{II} e^{-V^{II}}D_-e^{V^{II}}+D_+\bar{\Psi}^{II} e^{V^{II}}\bar{D}_-e^{-V^{II}}\right) \nn\\
&-\frac{1}{2e^2}\bar{\Sigma}_0\Sigma_0\bigg)+\frac{1}{2}\left(-\int d^2\tilde{\theta}t\Sigma_0+c.c\right).\nn
\end{align}

The trace term for one $SU(2)$ factor in (\ref{Ldual1}) can be worked as

\begin{align}
\tilde{L}&\supset-\frac{1}{2}\int d^4\theta\text{Tr}\left(\bar{D}_+\Psi^I e^{-V}D_-e^V+D_+\bar{\Psi^I} e^{V^I}\bar{D}_-e^{-V^I}\right),\\
&=-\frac{1}{2}\int d^4\theta\text{Tr}\left(\bar{D}_+\Psi^I\hat{n}D_-|V^I|-D_+\bar{\Psi^I}\hat{n}\bar{D}_-|V^I| \right),\notag\\
&=-\int d^4\theta\left( D_-\bar{D}_+\Psi^I_a\hat{n}_a|V^I|+D_+\bar{D}_-\bar{\Psi^I}_a\hat{n}_a|V^I|\right),\nn \\
&=-\int d^4\theta\left( X_a\hat{n}_a|V^I|+\bar{X}_a\hat{n}_a|V^I|\right). \nn
\end{align}

Substituting the vector superfield (\ref{VIexp}) in previous expression we get
\begin{align}
\tilde{L}&\supset-\int d^4\theta\left( X_a\hat{n}_a|V^I|+c.c\right)=-\int d^4\theta\left[X_a\hat{n}_a\mathcal{K}(X_a,\bar{X}_a,\hat{n}_a)+c.c\right]\\
&-\frac{1}{2}\int d\theta^+\bar{\theta}^-\left(X_a\left(-\bar{D}_+D_-\text{Ln}(2|\Phi_2|^2)\frac{1}{2}\hat{n}_a-2Q^I\hat{n}_a\Sigma_0-Q^I\bar{D}_+\hat{n}D_-V_0\right)\right)+c.c,\notag\\
&=-\int d^4\theta\left[X_a\hat{n}_a\mathcal{K}(X_a,\bar{X}_a,\hat{n}_a)+c.c\right]+\frac{Q^I}{2}\int d\theta^+\bar{\theta}^-\left(2X_a\hat{n}_a\Sigma_0+X_a\bar{D}_+\hat{n}_aD_-V_0\right)+c.c.  \nn
\end{align}
As it was concluded previously the formula
applies for the second $SU(2)$ factor of the dualization by exchanging the index $I$ by $II$
and $\hat n_a$ by $\hat m_a$. The total dual Lagrangian is therefore given by
 
\begin{align}
\tilde{L}&=\int d^4\theta\Bigg\{\sqrt{({Z_1}^I)^2+({Z_2}^{I})^2+({Z_3}^{I})^2}-\left(X_{a}\hat{n}_{a}\mathcal{K}(X_{a},\bar{X}_{a},\hat{n}_{a})+c.c\right)\Bigg\}\label{dualL2}\\
&+Q_{1}\left(\int d\theta^+d\bar{\theta}^-\left[\left(\hat{n}_{a}X_{a}-\frac{t}{4Q^{I}}\right)\Sigma_0+\frac{X_{a}\bar{D}_+\hat{n}_{a}D_-V_0}{2}\right]+c.c\right)\notag\\
&+\int d^4\theta\Bigg\{\sqrt{({Z_1}^{II})^2+({Z_2}^{II})^2+({Z_3}^{II})^2}-\left(Y_{a}\hat{m}_{a}\mathcal{K}(Y_{a},\bar{Y}_{a},\hat{m}_{a})+c.c\right)\Bigg\}\notag\\
&+Q^{II}\left(\int d\theta^+d\bar{\theta}^-\left[\left(\hat{m}_{a}Y_{a}-\frac{t}{4Q^{II}}\right)\Sigma_0+\frac{Y_{a}\bar{D}_+\hat{m}_{a}D_-V_0}{2}\right]+c.c\right)\nn\\
&-\int d^4\theta\frac{1}{2e^2}\bar{\Sigma}_0\Sigma_0.\nn
\end{align}

\section{Dualization for an Abelian direction inside de non-Abelian group}

In this section we analyze the geometry of the dual model considering a dualization along a family of Abelian directions inside
of the non-Abelian gauged group $SU(2)\times SU(2)$. This is given by the constant vectors $(n_1,n_2,n_3)$
and $(m_1,m_2,m_3)$ such that $V^I=V^I_a T_a= |V^I| n_a T_a$ and  $V^{II}=V^I_a T_a= |V^{II}| m_a T_a$. The superfields $\hat n_a,\hat m_a$ are substituted by the constant values $\hat n_a=n_a$ and $\hat m_a=m_a$.

For the $SU(2)\times SU(2)$ symmetry case integrating the gauged vector fields $V^I$ and $V^{II}$ in (\ref{Lmaster}) the dual model gives two copies of the same Lagrangian:
\begin{eqnarray}
L_{dual,0}&=&\int d^4\theta \left(\sqrt{\sum_a( X_a  + \bar X_a)^2}+ \sqrt{\sum_a( Y_a  + \bar Y_a)^2}\right)\label{exT}\\
&+& \int d^4\theta  (X_a n_a \ln(\mathcal{K}(X_i,\bar X_i,n_j))+Y_a m_a \ln(\mathcal{K}(Y_i,\bar Y_i,m_j)))+c.c.\nn\\
&+&\frac{1}{2}\int d\bar \theta^-d\theta^+\left(X_a n_a-Y_a m_a-t/2\right) \Sigma_0+c.c.,\nn\\
&+&\int d\bar \theta^-d\theta^+\left(X_a \bar D_+n_a-Y_a \bar D_+m_a\right) D_-V_0+c.c.\nn \\
&-& \int d^4\theta\frac{1}{2 e^2}\bar{\Sigma}_0\Sigma_0.\nn
\end{eqnarray}
As it occurs for the Mirror symmetry map: The csf $\Phi_1,\Phi_2,\Phi_3,\Phi_4$ are exchanged by  twisted csf 
$X_a,Y_a,a=1,2,3$.  This gives rise to the potential
\begin{eqnarray}
U_{dual,0}=8 e^2 (t + 
   Q (-n_{1} X_{1} - n_{2}X_{2} - n_{3} X_{3} + m_{1} Y_1 + m_{2} Y_2 +    m_3 Y_3))\times \\
(\bar{t} + 
   Q (-n_{1} \bar X_{1} - n_{2}\bar X_{2} - n_{3} \bar X_{3} + m_{1} \bar Y_1 + m_{2} \bar Y_2 +    m_3 \bar Y_3)). \nn
      \end{eqnarray}
      The SUSY vacuum  is given by the following set of six real equations:
      \begin{eqnarray}
      (t + 
   Q (-n_{1} X_{1} - n_{2}X_{2} - n_{3} X_{3} + m_{1} Y_1 + m_{2} Y_2 +    m_3 Y_3))&=&0, \\
  t+Q(Y_a m_a-X_a n_a) &=&0. \nn
         \end{eqnarray}
      
      Taking into account the $SU(2)\times SU(2)$ gauge freedom to set $X_2+\bar X_2=X_3+\bar X_3=Y_2+\bar Y_2=Y_3+\bar Y_3=0$ \cite{CaboBizet:2017fzc}. This can be written as
      
         \begin{eqnarray}
   &&Q (-n^I_{1} Re(X_{1}) +m_{1}Re(Y_{1}) )=-t_1. \label{dualM}\\
 &&Q (-n_{1} Im(X_{1}) - n_{2}Im(X_{2}) -m_{3} Im(Y_{3}) + m_{1}Im( Y_1) + m_{2} Im(Y_2) +    m_3 Im(Y_3))=-t_2.
     \nonumber
         \end{eqnarray}
         
 We employ the quantum symmetries of (\ref{exT}) given by:       
                        \begin{eqnarray}
 Im(X_{a}) \rightarrow  Im(X_{a}) + \frac{2\pi r_{a}}{n_a}, r_a\in \mathbb{Z},\label{Qsym}\\
  Im(Y_{a}) \rightarrow  Im(Y_{a}) + \frac{2\pi s_{a}}{m_a}, s_a\in \mathbb{Z}. \nn
     \nonumber
         \end{eqnarray}
They arise due to the $\theta$ term symmetry of the theory, i.e. a theory with parameter $\theta$ is equivalent
to a theory with parameter $\theta+2 k\pi, k \in \mathbb{Z}$. Notice the periodicity of the fields, this
is due to the fact that the symmetry is also required to be a symmetry of the <superpotential. 

The SUSY vacuum of this model gives for the  dual model the geometry: $T^5\times \mathbb{R}$. 
Let us explain this fact. There are 12 real coordinates given by $Re(X_a),Im(X_a),Re(Y_a),Im(Y_a)$, with $a=1,2,3$.
There are conditions from fixing the gauge $Re(X_2)=Re(X_3)=Re(Y_2)=Re(Y_3)=0$, this leaves
8 real coordinates. Now due to the equation (\ref{dualM}) one can eliminate $Re(X_1)$ and $Im(X_1)$,
this leaves 6 real coordinates: $Im(X_2)$, $Im(X_3)$, $Re(Y_1)$, $Im(Y_1)$, $Im(Y_2)$, $Im(Y_3)$. Together
with the identifications (\ref{Qsym}) this makes the geometry of $T^5\times \mathbb{R}$.

Let us discuss the $U(1)_A$ and $U(1)_V$ R-Symmetries for this model, specially the
axial R-symmetry. The $U(1)_V$ is preserved in the dual model, and the  dual scalar fields should not transform under them. Thus it should not play a role in the geometry of the dual model. They could affect the
dual manifold, because the fields $X_a$ and $Y_a$ can be charged with respect to them. 
The equations of motion lead to the fact that the dual fields have $0$ charge
under both R-charges \cite{cabosantos19}. However the expansion of the $U(1)_A$ currents imply that
the superfields could have a shift transformation with respect to $U(1)_A$.
The symmetry is anomaly free as the departure model satisfies the Calabi-Yau condition
$\sum_i Q_i=0$ \cite{Witten:1993yc}. To know the details of this transformation from $X_a,Y_a$ the OPE of the currents with
the fields needs to be computed. However we can describe some facts of this symmetry.
Looking at the twisted superfield action
$\widetilde{W}=((X_a n_a-Y_a m_a-t/2)\Sigma_0+ e^{-X_a n_a}+e^{-Y_a n_a})$, any
symmetry surviving in the dual theory should be a symmetry of it, or in case
it is an R-symmetry it should give the phase transformation $\widetilde{W}\rightarrow e^{2 i \alpha} W$ when $\theta^+\rightarrow e^{-i \alpha}\theta^+,\bar{\theta}^-\rightarrow e^{-i \alpha}\bar{\theta}^-$ such that $\int d\theta^+\bar{\theta}^-\rightarrow e^{-2 i \alpha}\int d\theta^+\bar{\theta}^-$. First the fields $X_a$ and $Y_a$ should have the exact same transformation property, because the equations of motion coincide \cite{cabosantos19}. 
The instanton terms should transform as $ e^{-X_a n_a}\rightarrow e^{-2 i \alpha}  e^{-X_a n_a}$ and  $ e^{-Y_a m_a}\rightarrow e^{-2 i \alpha}  e^{-Y_a m_a}$, so one should have the identifications $X_a n_a\rightarrow X_a n_a-2 i\alpha$ and $Y_a m_a\rightarrow X_a m_a-2 i\alpha$.
The consequence of this symmetry needs to be analyzed in further detail, by performing the
corresponding computations of the operator
product expansions of the $U(1)_A$ currents with the dual fields. This is the subject of future work.

\section{Dualization in an Abelian direction with instanton corrections}

In this section we implement an Ansatz for the instanton corrections of the dual theory
in the case of a generic Abelian direction inside the non Abelian duality. They can be interpreted
as a family of Abelian directions parametrized by te coefficients $n_a$ and $m_a$.
First we study the instanton for the case when one has one $SU(2)$
global symmetry. Then we focus in our case of interest, the global
symmetry group $SU(2)\times SU(2)$. The SUSY vacua in the dual theory are given by the derivatives of the twisted superpotential $\widetilde{W}$ equal to zero,
which gives rise to the condition on the scalar potential $U=0$.

Let us work out the corrections for the case a GLSM with a single SU(2) global symmetry.
The dual Lagrangian incorporating the Ansatz reads
\begin{eqnarray}
L_{dual}&=&\int d^4\theta \left(\sqrt{\sum_a( X_a  + \bar X_a)^2}\right)\label{exT}\\
&+& \int d^4\theta  (X_a \hat n_a \ln(\mathcal{K}(X_i,\bar X_i,n_j)))+c.c.\nn\\
&+&\frac{1}{2}\int d\bar \theta^-d\theta^+\left(X_a n_a-t/2\right) \Sigma_0+c.c.,\nn\\
&+&\int d\bar \theta^-d\theta^+\left(X_a \bar D_+n_a\right) D_-V_0+c.c.\nn \\
&-& \int d^4\theta\frac{1}{2 e^2}\bar{\Sigma}_0\Sigma_0,\nn\\
&+& \mu \int d^2\tilde\theta(e^{-X_a n_a})+\mu  \int d^2\bar{\tilde \theta} (e^{-\bar X_a n_a}).\nn
\end{eqnarray}

Integrating $\sigma_0$ we obtain the condition $X_an_a=t/2$, this implies a constant twisted superpotential,
and a scalar potential $U=0$ in the effective theory. Therefore there are not extra constraints and even including the instanton
corrections the dual geometry is preserved to be also a 2D torus $T^2$
as analyzed in \cite{CaboBizet:2017fzc}. In this case
the $U(1)_A$ symmetry is broken to a subgroup, given by the $\theta-$term
symmetry, so that it doesn't restricts more the resulting space.

For the case of the global symmetry $SU(2)\times SU(2)$ for the instanton corrections one obtains the following Lagrangian:
 \begin{eqnarray}
L_{dual}&=&L_{dual,0}+ \mu \int d^2\tilde\theta(e^{-X_a n_a}+e^{-Y_a m_a})+\mu  \int d^2\bar{\tilde \theta} (e^{-\bar X_a n_a}+e^{-\bar Y_a m_a}).
  \end{eqnarray}
  
 The instanton corrections can be obtained as an Ansatz by comparing the effective twisted superpotential for $\Sigma_0$, $W_{eff}$ in both theories. The instanton
 term should lead to an $W_{eff}$ that coincides with the one calculated at one loop in the original theory \cite{Witten:1993yc}:
 \begin{eqnarray}
\widetilde{W}_{eff}(\Sigma_0)&=&\sum_i \Sigma_0 Q_i \left(\ln\left(\frac{Q_i\Sigma_0}{\mu}\right)-1\right)-t\Sigma_0.
  \end{eqnarray}
  For the present case (the resolved conifold GLSM) we have
   \begin{eqnarray}
\widetilde{W}_{eff}(\Sigma_0)&=&\sum_i \Sigma_0 Q_i \left(\ln\left(\frac{Q_i\Sigma_0}{\mu}\right)\right)-t\Sigma_0,\\
&=&(-2i\pi-t)\Sigma_0. \label{Weff1}
  \end{eqnarray}
  This occurs because the charges of the fields are $Q_1=Q_2=1$ and $Q_3=Q_4=-1$. That the instanton Ansatz is correct can be seen by
  considering the equations of motion
     \begin{eqnarray}
\frac{\partial \widetilde{W}}{\partial X_a}= -n_a e^{-X_a n_a}\mu+2 n_a \Sigma_0=0,\\
\frac{\partial \widetilde{W}}{\partial Y_a}=  -m_a e^{-Y_a m_a}\mu-2 m_a \Sigma_0=0.
  \end{eqnarray}
  Obtaining as a solution $n_a X_a=-\ln(2\Sigma_0/\mu)$ and $m_a Y_a=-\ln(-2\Sigma_0/\mu)$,
  giving $\widetilde{W}=(-2i((2k+1)\pi)-t)$. This superpotential coincides with the previous (\ref{Weff1})
  taking into account the $\theta$ symmetries. The contribution of these instanton corrections to the scalar potential can be written as:
 \begin{eqnarray}
U_{dual}&=&g^{\mu \bar\nu} \partial_{\mu} \widetilde{W}\partial_{\bar\nu} \bar{\widetilde{W}}.\label{Udual}
  \end{eqnarray}

Let us integrate first the field strength $\sigma_0$. The variation of the twisted superpotential with respect to $\Sigma_0$ gives
 \begin{eqnarray}
\left(2 X_a n_a-2 Y_a m_a-t\right) =0.
  \end{eqnarray}

Then we have 
 \begin{eqnarray} 
 \widetilde W=e^{(-X_a n_a)} + e^{(-t/2-X_a n_a)}. 
\end{eqnarray}
Let us study the vacua space 
$\partial_{X_a}\widetilde{W}=0$, it gives the solution $e^{-t/2}=-1$, which is satisfied by $t/2=(2k+1)\pi i, k\in \mathbb{Z}$.
Implying that $r=0$ and the dual is restricted to the singular conifold. Let us notice an interesting fact
this is the same result obtained if one analyzes the effective theory for $U(1)$ field strength in the original
theory, which is given by the twisted superpotential (\ref{Weff1}). Integrating $\Sigma_0$ there one obtains
the same restriction for the parameter $t$.

\subsubsection{Instanton corrections in the full dual theory}
In order to compute the scalar potential we use the formulae $(3.7)$ in the reference  \cite{cabosantos19}. Considering only the bosonic fields and integrating out the auxiliary fields, it reduces to (\ref{Udual}) $U_{dual,0}=g^{\mu \bar\nu} (\partial_\mu \widetilde W)(\partial_{\bar\nu} \bar{\widetilde W})$, 
where $\tilde W= (X_a n_a -Y_a m_a -\frac{1}{2}t )\sigma_0 + \mu (e^{-X_a n_a} +e^{-Y_a m_a})$. We must find the critical points of this expression, i.e. compute $\partial_\mu \tilde W=0$ then we obtain the expressions:

\begin{eqnarray}
\frac{\partial \widetilde{W}}{\partial \sigma_0}= X_a n_a - Y_am_a -\frac{1}{2}t =0, \\
\frac{\partial \widetilde{W}}{\partial X_b}= n_b\sigma_0 - \mu n_b e^{-X_a n_a}=0, \\
\frac{\partial \widetilde{W}}{\partial Y_b}=- m_b\sigma_0 -\mu m_b e^{-Y_a m_a}=0,
\end{eqnarray}
from these formulae we can solve $\sigma_0$ and express $Y_a$'s in terms of $X_a$'s, thus we obtain the expression

\begin{equation}
e^{-X_a n_a}=-e^{-X_a n_a +\frac{1}{2}t}.
\end{equation}
It is worth to mention that this expression can be also obtained substituting the first equation into to the superpotential and set in it to zero. This equation can be solved by $t/2=(2k-1)\pi i, k\in \mathbb{Z}$. Which is the same result obtained before
by considering the effective twisted superpotential for the twisted chiral superfields. This also coincides
with the observation in the effective superpotential for 
$\Sigma_0$ in the original model, which upon integration gives rise to the 
same constraint for the $t$ parameter.


\section{Dualization on a non-Abelian direction, with a semichiral vector Superfield}

\label{semichiral}

In this section we study the dualization considering a truly non-Abelian duality. However the full $SU(2)\times SU(2)$ non-Abelian duality  is simplified by chosing a semi-chiral representation for the superfields $n_a=V^I_a/V^I$ and $m_a=V^I_a/V^I$.
This allows to determine the dual model with
more ease.

In order to calculate the contributions to the scalar potential we have to write the field $\hat{n}_a$ in components. A general real superfield in terms of the $y$ variable is given by:
\begin{align}
w^\dagger(\bar{y}^\mu,\theta,\bar{\theta})&=C(\bar{y})+i(\theta^-\gamma^+(\bar{y})-\theta^+\gamma^-(\bar{y}))-i(\bar{\theta}^+\bar{\gamma}^-(\bar{y})-\bar{\theta}^-\bar{\gamma}^+(\bar{y}))\notag\\
&+i\theta^-\theta^+(M(\bar{y})+iN(\bar{y}))+i\bar{\theta}^-\bar{\theta}^+(M(\bar{y})-iN(\bar{y}))+\theta^+\bar{\theta}^-(w_1(\bar{y})-iw_2(\bar{y}))\notag\\
&+\theta^-\bar{\theta}^+(w_1(y)+iw_2(\bar{y}))-\theta^+\bar{\theta}^+(w_0(\bar{y})-w_3(\bar{y}))-\theta^-\bar{\theta}^-(w_0(\bar{y})+w_3(\bar{y}))\notag\\
&+2\bar{\theta}^+\bar{\theta}^-(\theta^-\lambda^+(\bar{y})-\theta^+\lambda^-(\bar{y}))+2\theta^-\theta^+(\bar{\theta}^+\bar{\lambda}^-(\bar{y})-\bar{\theta}^-\bar{\lambda}^+(\bar{y}))\notag\\&+4\theta^-\theta^+\bar{\theta}^+\bar{\theta}^-D(\bar{y}), \label{wreal}
\end{align}
where the variables are given as $y^\mu=x^\mu+i\theta^\alpha\sigma^\mu_{\alpha\dot{\alpha}}\bar{\theta}^{\dot{\alpha}}$ and $\bar{y}^\mu=x^\mu-i\theta^\alpha\sigma^\mu_{\alpha\dot{\alpha}}\bar{\theta}^{\dot{\alpha}}$, or more expicitly as $y^\mu=x^\mu-i(\pm\theta^+\bar{\theta}^++\theta^-\bar{\theta}^-)$ and $\bar{y}^\mu=\bar{x}^\mu+i(\pm\theta^+\bar{\theta}^++\theta^-\bar{\theta}^-)$.\\

In \cite{CaboBizet:2017fzc} the procedure to obtain this semi-chiral superfield which is also real was carried out. One takes (\ref{wreal}) and imposes the semi-chirality condition, here we summarize this results. In addition the semi-chiral condition $D_- \hat{n}_a=0$ implies that $\hat{n}_a^\dagger$ has no explicit dependence on fermionic coordinate $\theta^-$, so that it can be written as
\begin{eqnarray}
\hat{n}_a&=&C(\bar{y})-i\theta^+\gamma^-(\bar{y})-i(\bar{\theta}^+\bar{\gamma}^-(\bar{y})-\bar{\theta}^-\bar{\gamma}^+(\bar{y}))\notag\\
&+&i\bar{\theta}^-\bar{\theta}^+(M(\bar{y})-iN(\bar{y}))+\theta^+\bar{\theta}^-(w_1(\bar{y})-iw_2(\bar{y}))\notag\\
&-&\theta^+\bar{\theta}^+(w_0(\bar{y})-w_3(\bar{y}))+2\bar{\theta}^+\bar{\theta}^-\theta^+\lambda^-(\bar{y})\\
&=&C(x)+i\theta^+\bar{\theta}^+(\partial_0-\partial_3)C(x)+i\theta^-\bar{\theta}^-(\partial_0+\partial_3)C(x)\notag\\
&-&\theta^+\bar{\theta}^+\theta^-\bar{\theta}^-(\partial_0^2-\partial_3^2)C(x)-i\theta^+\gamma^-(x)+\theta^+\theta^-\bar{\theta}^-(\partial_0+\partial_3)\gamma^-(x)\notag\\
&-&i(\bar{\theta}^+\bar{\gamma}^-(x)-\bar{\theta}^-\bar{\gamma}^+(x))+\bar{\theta}^+\theta^-\bar{\theta}^-(\partial_0+\partial_3)\bar{\gamma}^-(x)-\bar{\theta}^-\theta^+\bar{\theta}^+(\partial_0-\partial_3)\bar{\gamma}^+(x)\notag\\
&+&i\bar{\theta}^-\bar{\theta}^+(M(x)-iN(x))+\theta^+\bar{\theta}^-(w_1(x)-iw_2(x))\notag\\
&-&\theta^+\bar{\theta}^+(w_0(x)-w_3(x))-i\theta^+\bar{\theta}^+\theta^-\bar{\theta}^-(\partial_0+\partial_3)(w_0(x)-w_3(x))\notag\\
&+&2\bar{\theta}^+\bar{\theta}^-\theta^-\lambda^-(x).
\end{eqnarray}

The reality condition meas that $\hat{n}_a(x)=\hat{n}_a^\dagger$. The expansion for it is given by

\begin{eqnarray}
\hat{n}_a^\dagger&=&C(x)-i\theta^+\bar{\theta}^+(\partial_0-\partial_3)C(x)-i\theta^-\bar{\theta}^-(\partial_0+\partial_3)C(x)\notag\\
&-&\theta^+\bar{\theta}^+\theta^-\bar{\theta}^-(\partial_0^2-\partial_3^2)C(x)-i\bar{\theta}^+\bar{\gamma}^-(x)+\bar{\theta}^+\bar{\theta}^-\theta^-(\partial_0+\partial_3)\bar{\gamma}^-(x)\notag\\
&-&i(\theta^+\gamma^-(x)-\theta^-\gamma^+(x))+\theta^+\bar{\theta}^-\theta^-(\partial_0+\partial_3)\gamma^-(x)-\theta^-\bar{\theta}^+\theta^+(\partial_0-\partial_3)\gamma^+(x)\notag\\
&+&i\theta^-\theta^+(M(x)+iN(x))-\bar{\theta}^+\theta^-(w_1(x)+iw_2(x))\notag\\
&-&\theta^+\bar{\theta}^+(w_0(x)-w_3(x))+i\theta^+\bar{\theta}^+\theta^-\bar{\theta}^-(\partial_0+\partial_3)(w_0(x)-w_3(x))\notag\\
&+&2\theta^+\theta^-\bar{\theta}^-\bar{\lambda}^-(x).
\end{eqnarray}

Due to the reality condition one obtains the restricted expansion \cite{CaboBizet:2017fzc}:

\begin{eqnarray}
\hat{n}_a&=&n_a(x)+2\theta^+\bar{\theta}^+\theta^-\bar{\theta}^-(\partial_0^2-\partial_3^2)n_a(x)+i\theta^+\gamma_+^a(x)+\theta^+\bar{\theta}^+(w^a_0(x)-w^a_3(x))\notag\\
&+&i\bar{\theta}^+\bar{\gamma}_+^a(x)-\theta^+\theta^-\bar{\theta}^-(\partial_0+\partial_3)\gamma^a_+-\bar{\theta}^+\bar\theta^-\theta^-(\partial_0+\partial_3)\bar{\gamma}^a_+.
\end{eqnarray}

Let us compute the contributions to the scalar potential of the dual Lagrangian in (\ref{dualL2}) as:
\begin{align}
\hat{n}_aX_a\Sigma_0=\sqrt{2}Q\left(n_a\bar{\Delta}_aD_0-2n_a\bar{\sigma}_0G_a\right)+i\sqrt{2}n_a\bar{\Delta}_av_{03}+c.c.
\end{align}
Other contribution comes from
\begin{align}
\frac{Q}{2}X_a\bar{D}_+\hat{n}_aD_-V_0=\frac{Q}{2}
\sqrt{2}\bar{\Delta}_a(v_0+v_3)\omega_{03}^a+c.c.
\end{align}

The component expansion of the integral with the terms of the semichiral real superfield $\hat{n}_a$ is given by:

\begin{align}
\hat{n}_aX_a\Sigma_0=&n_a\big[2\lambda^+_0\left(-2i\bar{\rho}^-_a+(\partial_0-\partial_3)w^+_a\right)+2\bar{\lambda}_0^-\left(-2i\mu_a^++\right(\partial_0+\partial_3)\bar{\xi}_a^-)\big]\notag\\
&+2\sqrt{2}\gamma^-_a\left(-\lambda^+_0\bar{\Delta}_a+\bar{\sigma}_0\left(i\mu_a^+-\frac{1}{2}\left(\partial_0+\partial_3\right)\bar{\xi}_a^-\right)\right)\notag\\
&+\sqrt{2}n_a\big[\bar{\Delta}_a\big(D_0+iv_{03}\big)+\bar{\sigma}_0\big(-2\sqrt{2}J_a+2ib_{03}^a-\big(\partial_0^2-\partial_3^2\big)\varphi_a\big)\big].
\end{align}  

So its contribution to the scalar potential is
\begin{equation}
\sqrt{2}Q\left(n_a\bar{\Delta}_aD_0-2\sqrt{2}n_a\bar{\sigma}_0G_a\right).
\end{equation}

The other new contribution comes from
\begin{align}
X_a\bar{D}_+\hat{n}_aD_-V_0=&-\big\{\gamma^-_a\big[(v_0+v_3)\big(2i\bar{\rho}_a^--(\partial_0-\partial_3)w^+_a\big)+2\sqrt{2}\bar{\lambda}^+_a\bar{\Delta}_a\big]\notag\\
&-\sqrt{2}\bar{\Delta}_a(v_0+v_3)\omega_{03}^a\big\},
\end{align}
such that for the potential one has:
\begin{equation}
-\frac{Q}{2}\sqrt{2}\bar{\Delta}_a(v_0+v_3)\omega_{03}^a,
\end{equation}
with $v_{03}\equiv\partial_0v_3-\partial_3v_0$, $b_{03}\equiv\partial_0b_3-\partial_3b_0$ and $\omega_{03}\equiv\omega_0-\omega_3$.

Let us clarify that through out this section a notation will be employed to denote the 
two $SU(2)$ subgroups of the duality given by $1,2$.
Such that the sub-indices $a_1,b_,i_1,\bar j_1$ correspond to the first subgroup and the sub-indices $a_2,b_2, i_2,\bar j_2$ to the second.
We will also denote the charges as $Q_1\equiv Q^I$
and $Q_2\equiv Q^{II}$. So the previous formulae developed for a semi-chiral superfield $\hat n_a$ will have a sub-index with a $1$ or $2$.

Finally the full scalar potential in the dual model is given by
\begin{align}
\tilde{U}&=K_{i_1\bar{j}_1}G^{i_1}\bar{G}^{\bar{j}_1}+K_{i_2\bar{j}_2}G^{i_2}\bar{G}^{\bar{j}_2}-\frac{D_0^2}{8e^2}+\sqrt{2}D_0\left(Q_1n_{a_1}(\bar{\Delta}_{a_1}+\Delta_{a_1})+Q_2n_{a_2}(\bar{\Delta}_{a_2}+\Delta_{a_2})\right)\notag\\
&-2\sqrt{2}\left(Q_1n_{a}(\bar{\sigma}_0G_{a_1}+\sigma_0\bar{G}_{a_1})+Q_2n_{a_2}(\bar{\sigma}_0G_{a_2}+\sigma_0\bar{G}_{a_2})\right)\notag\\
&+\frac{\sqrt{2}}{2}(v_0+v_3)\left(Q_1\omega_{03}^{a_1}(\bar{\Delta}_{a_1}+\Delta_{a_1})+Q_2\omega_{03}^{a_2}(\bar{\Delta}_{a_2}+\Delta_{a_2})\right)-\frac{1}{2}(t+\bar{t})D_0,
\end{align}
with the indices $a=1,2,3$, $i_1,j_1=1,2,3,i_2,j_2=1,2,3$  denoting the $SU(2)$ generators.  The fields $x_{1a}$ and $x_{2a}$ are the scalar components of the twisted superfields $X_a$ and $Y_a$ respectively. At the end of the calculations we will homogenize the notation with the one employed in previous sections. Eliminating the $U(1)$ auxiliary field $D_0$ with the equations of motion we have
\begin{eqnarray}
D_0=2e^2\left(2\sqrt{2}\left(Q_1n_{a_1}x_{a_1}+Q_2n_{a_2}x_{a_2}\right)-t_R\right).
\end{eqnarray}
Then the potential reads
\begin{align}
\tilde{U}&=K_{i_1\bar{j}_1}G^{i_1}\bar{G}^{\bar{j}_1}+K_{i_2\bar{j}_2}G^{i_2}\bar{G}^{\bar{j}_2}+2e^2(2\sqrt{2}\left(Q_1n_{a_1}x_{a_1}+Q_2n_{a_2}x_{a_2})-t_R\right)^2\notag\\
&-\sqrt{2}(v_0+v_3)\left(Q_1x_{a_1}\omega^{a_1}_{03}+Q_2x_{a_2}\omega^{a_2}_{03}\right)\notag\\
&-2\sqrt{2}\left(Q_1n_{a_1}(\bar{\sigma}_0G^{a_1}+\sigma_0\bar{G}^{a_1})+Q_2n_{a_2}(\bar{\sigma}_0G^{a_2}+\sigma_0\bar{G}^{a_2})\right).
\end{align}

Integrating now the auxiliary fields $G$ we obtain
the equations of motion
\begin{eqnarray}
K_{a_1\bar{b}_1}\bar{G}^{\bar{b}_1}&=&2\sqrt{2}Q_1n_{a_1}\bar{\sigma}_0,\\
K_{a_2\bar{b}_2}\bar{G}^{\bar{b}_2}&=&2\sqrt{2}Q_2n_{a_2}\bar{\sigma}_0,\nonumber\\
K_{a_1\bar{b}_1}G^{a_1}&=&2\sqrt{2}Q_1n_{b_1}\sigma_0,\nonumber\\
K_{a_2\bar{b}_2}G^{a_2}&=&2\sqrt{2}Q_2n_{b_2}\sigma_0. \nonumber
\end{eqnarray}
Writing these expressions back in the potential we have:


\begin{align}
\tilde{U}&=-8|\sigma_0|^2({K^{-1}}^{a_1\bar{b}_1} Q_1^2n_{a_1}n_{b_1}+{K^{-1}}^{a_2\bar{b}_2}Q_2^2n_{a_2}n_{b_2})\\
&+2e^2(2\sqrt{2}\left(Q_1n_{1a}x_{a_1}+Q_2n_{2a}x_{a_2})-t_R\right)^2\notag\\
&-\sqrt{2}(v_0+v_3)\left(Q_1x_{a_1}\omega^{a_1}_{03}+Q_2x_{a_2}\omega^{a_2}_{03}\right). \nn
\end{align}
We focus on the branch $|\sigma_0|=0$ of supersymmetric vacua with $U=0$:
\begin{align}
\tilde{U}&=2e^2(\sqrt{2}\left(Q_1n_{1a}x_{1a}+Q_2n_{2a}x_{2a})-t_R\right)^2-\sqrt{2}(v_0+v_3)\left(Q_1x_{1a}\omega^{1a}_{03}+Q_2x_{2a}\omega^{2a}_{03}\right).
\end{align}
The effective scalar potential is obtained if we add to the previous potential the interaction term for the vector field which contains $v_{03}$:
\begin{align}
\tilde{U}_{eff}&=2e^2(\sqrt{2}\left(Q_1n_{1a}x_{1a}+Q_2n_{2a}x_{2a})-t_R\right)^2+\sqrt{2}(v_0+v_3)\left(Q_1x_{1a}\omega^{1a}_{03}+Q_2x_{2a}\omega^{2a}_{03}\right)\notag\\
&-\frac{v_{03}^2}{8e^2}+\frac{iv_{03}}{2}\big[2\sqrt{2}n_{1a}Q_1(\bar{\Delta}_{a1}-\Delta_{a1})+2\sqrt{2}n_{a2}Q_2(\bar{\Delta}_{a2}-\Delta_{a2})+t-\bar{t}\big].
\end{align}
which can be simplified to:
\begin{align}
\tilde{U}_{eff}&=2e^2(\sqrt{2}\left(Q_1n_{1a}x_{1a}+Q_2n_{2a}x_{2a})+t_R\right)^2+\sqrt{2}(v_0+v_3)\left(Q_1x_{1a}\omega^{1a}_{03}+Q_2x_{2a}\omega^{2a}_{03}\right)\notag\\
&-\frac{v_{03}^2}{8e^2}+\frac{iv_{03}}{2}\big[4\sqrt{2}n_{1a}Q_1i\text{Im}(x_{1a})+4\sqrt{2}n_{2a}Q_2i\text{Im}(x_{2a})+2it_I\big].
\end{align}
Let us integrate the component of the gauge field $v_{03}$ to obtain
\begin{eqnarray}
v_{03}=2e^2i\left[2\sqrt{2}in_{1a}Q_1\text{Im}(x_{1a})+2\sqrt{2}in_{2a}Q_2\text{Im}(x_{2a})+2it_I\right].
\end{eqnarray}
Thus the effective potential reads:
\begin{eqnarray}
U_{eff}&=&2e^2\Bigg[\bigg(t_R-\left(n_{1a}\text{Re}(x_{1a})-n_{2a}\text{Re}(x_{2a})\right)\bigg)^2+\bigg(t_I-\big(n_{1a}\text{Im}(x_{1a})-n_{2a}\text{Im}(x_{2a})\big)\bigg)^2\notag\\
&+&(v_0+v_3)\left(\text{Re}(x_{1a})\omega^{1a}_{03}-\text{Re}(x_{2a})\omega^{2a}_{03}\right).
\end{eqnarray}

We will employ now the gauge $X_{12}+\bar{X}_{12}=0$, $X_{13}+\bar{X}_{13}=0$, $X_{22}+\bar{X}_{22}=0$, $X_{23}+\bar{X}_{23}=0$. The locus of vacua is given by setting the derivatives of $U$ to zero, such that:

\begin{align}
\text{Re}(x_{11})&=-\frac{w_1}{4e^2n_1^2}(v_0+v_3)+\frac{m_1\text{Re}(x_{21})}{n_1}+\frac{t_1}{n_1}, \label{EQ1}\\
\text{Im}(x_{11})&=\frac{-n_2\text{Im}(x_{12})-n_3\text{Im}(x_{12})+m_1\text{Im}(x_{21})+m_2\text{Im}(x_{22})+m_3\text{Im}(x_{23})}{n_1}+\frac{t_2}{n_1}\label{EQ2}\\
n_{1}\omega_1&=m_{1}w_1. \nn
\end{align}
Substituting the constant $A_1=\frac{w_1}{4e^2}(v_0+v_3)$, we have:
\begin{align}
n_1\text{Re}(x_{11})-m_1\text{Re}(x_{21})&=t_1-\frac{A_1}{n_1^2},\\
n_a\text{Im}(x_{1a})-m_a\text{Im}(x_{2a})&=t_2.
\end{align}
Therefore, out of the twelve coordinates, we can eliminate six coordinates, four with the gauge and two with equations (\ref{EQ1}) and (\ref{EQ2}). This leaves us with six coordinates $\text{Re}(y_1), \text{Im}(x_2), \text{Im}(x_3), \text{Im}(y_1), \text{Im}(y_2),\ \text{and}\ \text{Im}(y_3)$.\\

At the quantum level a symmetry comes from the periodicity of $t$, where in the dual Largangian it manifests as $x_a\rightarrow x_a+\frac{2\pi i k_a}{2n_a Q}$, with $k_a\in \mathbb Z$. Therefore, this identification between the imaginary parts, $ \text{Im}(x_2),\ \text{Im}(x_3),\ \text{Im}(y_1),\ \text{Im}(y_2),\ \text{and}\ \text{Im}(y_3)$ give us a five dimensional torus $T^5$. Thus the vacuum of the space of the dual theory is also given by $T^5\times \mathbb{R}$.\\

The Hessian matrix's eigenvalues and eigenvectors show two positive eigenvalues corresponding the growing directions of the potential. The number of zero values equal to zero are the real dimensions of the vacua space and represent the directions where the potential is flat.

\begin{table}[htp]
\caption{Table representing the eigenvalues and their associated eigenvector for the Hessian matrix. The entrances correspond to the directions $\text{Re}(y_1), \text{Im}(x_2), \text{Im}(x_3), \text{Im}(y_1), \text{Im}(y_2),\ \text{and}\ \text{Im}(y_3)$.}
\begin{center}
\begin{tabular}{|c|c|}
\hline
Eigenvalues & Eigenvectors\\
\hline
$0$ & ($0,\frac{m_{3}}{n_{1}}, 0,0,0,0,0,1$)\\
\hline
$0$ & ($0,\frac{m_{2}}{n_{1}}, 0,0,0,0,1,0$)\\
\hline
$0$ & ($0,\frac{m_{1}}{n_{1}}, 0,0,0,1,0,0$)\\
\hline
$0$ & ($\frac{m_{1}}{n_{1}},0, 0,0,1,0,0,0$)\\
\hline
$0$ & ($0,-\frac{n_{3}}{n_{1}}, 0,1,0,0,0,0$)\\
\hline
$0$ & ($0,-\frac{n_{2}}{n_{1}}, 1,0,0,0,0,0$)\\
\hline
$4e^2(m_{1}^2+n_{1}^2)$ & ($-\frac{n_{1}}{m_{1}},0, 0,1,0,0,0,0$)\\
\hline
$4e^2(m_{1}^2+m_{2}^2+m_{3}^2+n_{1}^2+n_{2}^2+n_{3}^2)$ & ($0,-\frac{n_{1}}{m_{3}},-\frac{n_{2}}{m_{3}},-\frac{n_{3}}{m_{3}},0,\frac{m_{1}}{m_{3}},\frac{m_{2}}{m_{3}},1$)\\
\hline
\end{tabular}
\end{center}
\label{autovalNorma}
\end{table}%

Let us now look at the case of the geometry of the space of supersymmetric vacua without fixing the gauge. In this situation, the solution of the supersymmetric vacua is given by:
\begin{align}
\text{Re}(x_{11})&=-\frac{(v _0+v_3)w_1}{4e^{2}n_1^{2}}+\frac{t_1-n_2\text{Re}(x_{12})-n_3\text{Re}(x_{13})+m_1\text{Re}(x_{21})+m_2\text{Re}(x_{22})+m_3\text{Re}(x_{23})}{n_1}\\
\text{Im}(x_{11})&=\frac{m_1\text{Im}(x_{21})+m_2\text{Im}(x_{22})+m_3\text{Im}(x_{23})-n_2 \text{Im}(x_{12})-n_3\text{Im}(x_{13})}{n_1}+\frac{t_2}{n_1}\\
n_a\omega_a&=m_aw_a
\end{align}
Substituting the constant $A_1$, we have:
\begin{align}
n_a\text{Re}(x_{1a})-m_a\text{Re}(x_{2a})&=t_1-\frac{A_1}{n_1},\\
n_a\text{Im}(x_{1a})-m_a\text{Im}(x_{2a})&=t_2.
\end{align}
Therefore, we can fix two coordinates, obtaining a 10-dimensional vacua space with the following real coordinates left unfixed: $\text{Re}(x_2)$, $\text{Re}(x_3)$, $\text{Re}(y_1)$, $\text{Re}(y_2)$, $\text{Re}(y_3)$, $\text{Im}(x_2)$,
$\text{Im}(x_3)$, $\text{Im}(y_1)$, $\text{Im}(y_2)$, and $\text{Im}(y_3)$. Using the $\theta$ quantum symmetry as in the previous case, which it applies to the imaginary parts, we obtain that the geometry of the vacua space is given by $T^5\times \mathbb{R}^5$.
A list of the eigenvalues and their associated eigenvectors for the Hessian are shown below:
\begin{table}[H]
\caption{Table representing the eigenvalues and their associated eigenvector for the Hessian matrix. The entrances correspond to the directions: $\text{Re}(x_2)$, $\text{Re}(x_3)$, $\text{Re}(y_1)$, $\text{Re}(y_2)$, $\text{Re}(y_3)$, $\text{Im}(x_2)$,
$\text{Im}(x_3)$, $\text{Im}(y_1)$, $\text{Im}(y_2)$, and $\text{Im}(y_3)$.}
\begin{center}
\begin{tabular}{|c|c|}
\hline
Eigenvalues & Eigenvectors\\
\hline
$0$ & ($0,0,0,\frac{m_{3}}{n_{1}}, 0,0,0,0,0,0,0,1$)\\
\hline
$0$ & ($0,0,0,\frac{m_{2}}{n_{1}}, 0,0,0,0,0,0,1,0$)\\
\hline
$0$ & ($0,0,0,\frac{m_{1}}{n_{1}}, 0,0,0,0,0,1,0,0$)\\
\hline
$0$ & ($\frac{m_{3}}{n_{1}},0,0,0,0,0, 0,0,1,0,0,0$)\\
\hline
$0$ & ($\frac{m_{2}}{n_{1}},0,0,0,0,0, 0,1,0,0,0,0$)\\
\hline
$0$ & ($\frac{m_{1}}{n_{1}},0,0,0,0,0, 1,0,0,0,0,0$)\\
\hline
$0$ & ($0,0,0,-\frac{n_{3}}{n_{1}},0, 1,0,0,0,0,0,0$)\\
\hline
$0$ & ($0,0,0,-\frac{n_{2}}{n_{1}},1, 0,0,0,0,0,0,0$)\\
\hline
$0$ & ($-\frac{n_{3}}{n_{1}},0, 1,0,0,0,0,0,0,0,0,0$)\\
\hline
$0$ & ($-\frac{n_{2}}{n_{1}}, 1,0,0,0,0,0,0,0,0,0,0$)\\
\hline
$4e^2(m_{1}^2+m_{2}^2+m_{3}^2+n_{1}^2+n_{2}^2+n_{3}^2)$ & ($0,0,0,-\frac{n_{1}}{m_{3}},-\frac{n_{2}}{m_{3}}, -\frac{n_{3}}{m_{3}},0,0,0,\frac{m_{1}}{m_{3}},\frac{m_{2}}{m_{3}},1$)\\
\hline
$4e^2(m_{1}^2+m_{2}^2+m_{3}^2+n_{1}^2+n_{2}^2+n_{3}^2)$ & ($-\frac{n_{1}}{m_{3}},-\frac{n_{2}}{m_{3}}, -\frac{n_{3}}{m_{3}},0,0,0,\frac{m_{1}}{m_{3}},\frac{m_{2}}{m_{3}},1,0,0,0$)\\
\hline
\end{tabular}
\end{center}
\label{default}
\end{table}%




The results of this section imply that a family of Abelian-dualities inside the non-Abelian
one can cast very important features of the dual model. It is not clear if this is the case for more generic 
situations, but in the first example of a non-compact CY manifold we have checked it.
This was also observed in \cite{CaboBizet:2017fzc} for the case of the $\mathbb{CP}^1$ dual, and here
it was made a stronger case for it, by considering in that example the instanton
corrections as well.  One would still have to determine the instanton corrections
for this non-Abelian duality and study the geometry including this contribution,
to check if the same result holds. 

\section{Conclusions and outlook}

In this work we have explored non-Abelian T-dualities in 2D (2,2) gauged linear sigma models giving rise to
target space string geometries. We have focused in the case of the GLSM leading to the
resolved conifold. The dual model geometry has been explored in different regimes, encountering
a relevant coincidence of the different analysis.

From all perspectives we have obtained that the target space geometry in the dual model is given by
$T^5\times \mathbb{R}$. This were three different study perspectives. First we have simplified
the duality equations by considering a generic Abelian direction inside the non Abelian
group $SU(2)\times SU(2)$. This direction was parametrized by the constant
coefficients $n_a=V^I_a/|V|$ and $m_a=V^{II}_a/|V|$ normalized as $n_an_a=m_am_a=1$. 
Second, we have determine the instanton corrections in the dual model,
by proposing an Ansatz that guarantees that the effective twisted superpotential
for the $U(1)$ vector field strength $\Sigma_0$ coincides in the original and in the dual model.
The $U(1)$ effective superpotential of the original model is a previous result and is computed employing one-loop contributions \cite{Witten:1993yc}.
Analyzing the theory with this superpotential we arrive to the condition that
the FI and $\theta$ term parameter $t$ is such that its real part is  restricted to be $r=0$
and its imaginary part to be $\theta=(2k+1)\phi, k \in \mathbb{Z}$. This was initially a surprise,
however it is in agreement with the analysis of
the equations of motion for the effective $U(1)$ theory given by the
twisted superpotential for $\Sigma_0$ in the original gauged linear sigma model, i.e. integrating $\Sigma_0$
in that theory one obtains the same condition for the $t$ parameter. We interpret this fact
as that the full duality only works for the model in which the $t$ parameter
is fixed, giving rise to the singular conifold. Finally  we considered that the coefficients
$n_a$ are also superfields, with space-time dependence, this consideration sets a generic direction in the non-Abelian group,
but for simplicity we imposed a semi-chiral condition on them $D_- n_a=0$. 
For this representation of the vector superfield $V=|V| n_a T_a$ the geometry of the target space in the dual model is also given by $T^5\times \mathbb{R}$. 
The instanton corrections for the full non-Abelian duality have still to be determined, but
this is the subject of future work.
There is an important observation, and this it is the fact that in order to determine the dual geometry,
a quantum symmetry present in the dual action was crucial. This is a subset of the $\theta$ term
symmetry of the original action, under which the dual superfields have periodic transformation properties \cite{Hori:2000kt}.
In the dual theory also the axial $U(1)_A$ R-symmetry is preserved, and we analyze this fact.
In other models the surviving part of it is connected to the $\theta$ term discrete symmetry. The implications of $U(1)_A$ in the dual model has not been fully explored, but we hope that in future work we will be able to clarify this connection.

In conclusion we have explored the non-Abelian duality for the resolved conifold gauged linear sigma 
model, finding that the dual geometry model (the locus of SUSY vacua) is given by a 6D flat space: $T^5\times \mathbb{R}$.
Furthermore this feature can be already seen by analyzing a generic direction for an Abelian dualization
inside of the full duality group $SU(2)\times SU(2)$. As future work we
need to study GLSMs with compact Calabi-Yau target spaces, and  
determine precisely the role of the instanton corrections
in this correspondence. We are also interested in finding the geometrical consequences of
these dualities in a wider range of examples. As well
as to determine the implications of these dualities for new geometrical correspondences between Calabi-Yau varieties \cite{Gu:2018fpm,Gu:2019rty,Gu:2020ana}.

\section{Acknowledgements}
We specially thank Leopoldo Pando Zayas, Aldo Mart\'{\i}nez Merino and Oscar Loaiza Brito for important ideas and discussions. We thank useful discussions and comments from Kentaro Hori, Matthias Gaberdiel, Hugo Garc\'{\i}a Compe\'an, Rodrigo D\'{\i}az Correa, Albrecht Klemm, Martin Rocek, Mauricio Romo, Jorge Gabriel Le\'on Bonilla, Octavio Obreg\'on and Cunrum Vafa. NGCB, RSS and YJS thank the projects CONACyT A1-S-37752 {\it ``Teorías efectivas de la teoría de cuerdas, y sus aplicaciones a física de partículas y cosmología"}  and CIIC 2021 DAIP 148/2021 {\it ``Exploración del paisaje de la teoría de cuerdas: geometría, dualidades y aprendizaje de máquina"}.  NCGB thanks the support of the Institute of Theoretical Physics, H\"onggerberg Campus, ETH Z\"urich, where this work was completed. YJS thanks the support of CONACyT, which provided him a PhD fellowship which allowed to develop this research. 
\newpage

\begin{appendices}

\section{Expressions for $SU(2)$ vector superfields}

Let us write some useful formulas for the $SU(2)$ case, that we will employ in the work. The Lagrangian term of the trace, for a single $SU(2)$ factor, lets say $V^I$ can be expressed like
\begin{eqnarray}
\text{Tr}\left(\left(\chi\tau+\tau\chi\right)\Delta V\right)&=&\chi_a\tau_b\Delta V_c\text{Tr}(\sigma_a\sigma_b\sigma_c)+\tau_a\chi_b\Delta V_c\text{Tr}(\sigma_a\sigma_b\sigma_c),\\
&=&4i\epsilon_{abc}\chi_a\tau_b\Delta V_c,\nn \\
&=&4i\left(\left(\chi_1\tau_2-\chi_2\tau_1\right)\Delta V_3+\left(\chi_3\tau_1-\chi_1\tau_3\right)\Delta V_2+\left(\chi_2\tau_3-\chi_3\tau_2\right)\Delta V_1\right),\nn
\end{eqnarray}
\\
while its adjoint conjugate reads  $4i\epsilon_{abc}\chi_a^\dagger\tau_b^\dagger\Delta V_c^\dagger$, with $\chi=\chi_a\sigma_a$, $\tau=\tau_a\sigma_a$ and $\Delta V=\Delta V_a\sigma_a$.
Also working with $\chi$ in the Wess-Zumino gauge we have
\begin{align}
\chi_a\sigma_a&\equiv e^{-V_a\sigma_a}D_-e^{V_a\sigma_a}=\left(1-V_a\sigma_a+\frac{1}{2}V_aV_b\sigma_a\sigma_b\right)D_-\left(1+V_a\sigma_a+\frac{1}{2}V_aV_b\sigma_a\sigma_b\right) \nn\\
&=\left(D_-V_a-i\epsilon_{dfa}V_dD_-V_f\right)\sigma_a, \nn
\end{align}
such that for example $\chi_1$ can be written as
\begin{eqnarray}
\chi_1&=&D_-V_1-i\epsilon_{df1}V_dD_-V_f,\\
&=&D_-V_1-i(V_2D_-V_3-V_3D_-V_2). \nn
\end{eqnarray}

\section{Another way of finding the equations of motion}
In this appendix section we present another derivation of the equations of motion, 
by considering a gauge covariant superfield variation.

\subsection{Variations $\Delta V$ and $\Delta V^\dagger$}
The variation can be expressed as:
\begin{eqnarray}
\Delta V\equiv e^{-V}\delta e^V\implies \Delta V^\dagger=\delta e^V e^{-V}=e^{V}e^{-V}\delta e^V e^{-V}=e^{V}\Delta V e^{-V}
\end{eqnarray}

Using the Wess-Zumino gauge we have:

\begin{align}
e^V=&1+V_a\sigma_a+\frac{1}{2}V_aV_b\sigma_a\sigma_b=1+V_a\sigma_a+\frac{1}{2}(V_aV_a+i\epsilon_{abc}V_aV_b\sigma_c),\\
=&1+V_a\sigma_a+\frac{1}{2}V^2, \hspace{0.1 cm} \text{with}  \hspace{0.2 cm} i\epsilon_{abc}V_aV_b\sigma_c=0.
\end{align}
One can expand the gauge invariant superfield variation:
\begin{align}
\Delta V&= (1-V_a\sigma_a+\frac{1}{2}V^2)\delta(1+V_b\sigma_b+\frac{1}{2}V^2)\notag\\
&=(1-V_a\sigma_a+\frac{1}{2}V^2)(\sigma_b+V_b)\delta V_b\notag\\
&=(\sigma_a+V_a)\delta V_a-V_a\delta V_a-i\epsilon_{abc}V_a\delta V_b\sigma_c\notag\\
&=\sigma_a\delta V_a-i\epsilon_{abc}V_a\delta V_b\sigma_c\notag\\
&=(\sigma_a+i\epsilon_{abc}V_b\sigma_c)\delta V_a.
\end{align}

\begin{align}
\Delta V^\dagger&=\delta(1+V_a\sigma_a+\frac{1}{2}V^2) (1-V_b\sigma_b+\frac{1}{2}V^2)\notag\\
&=(\sigma_a+V_a)\delta V_a(1-V_b\sigma_b+\frac{1}{2}V^2)\notag\\
&=(\sigma_a+V_a)\delta V_a-V_a\delta V_a-i\epsilon_{abc}\delta V_aV_b\sigma_c\notag\\
&=\sigma_a\delta V_a-i\epsilon_{abc}\delta V_a V_b\sigma_c\notag\\
&=(\sigma_a-i\epsilon_{abc} V_b\sigma_c)\delta V_a.
\end{align}

Therefore the equations of motion are

\begin{align}
\frac{\delta L}{\Delta V_c}=&e^{2QV_0}\bar{\Phi}_i(e^V\cdot\sigma_c)_{ij}\Phi_j-2i\epsilon_{abc}\bar{D}_+\Psi_aD_-V_b+2\bar{D}_+\Psi_aV_aD_-V_c-2\bar{D}_+\Psi_aV_cD_-V_a\notag\\-&D_-\bar{D}_+\Psi_c-2i\epsilon_{abc}D_+\bar{\Psi}_a\bar{D}_-V_b-2D_+\bar{\Psi}_aV_a\bar{D}_-V_c+2D_+\bar{\Psi}_aV_c\bar{D}_-V_a\notag\\-&D_+\bar{D}_-\bar{\Psi}_c.
\end{align}

Given the Weiss-Zumino gauge we can directly compute the $\delta e^V$ like

\begin{align}
\delta e^V=\delta(1+V_a\sigma_a+\frac{1}{2}V^2)=(\sigma_a+V_a)\delta V_a.
\end{align}

So the variation of the kinetic term will be

\begin{equation}
\delta\left(\bar{\Phi}_i (e^{2QV_0+V})_{ij}\Phi_j\right)=e^{2QV_0}\bar{\Phi}_i(\delta e^V)_{ij}\Phi_j=e^{2QV_0}\bar{\Phi}_i(\sigma_a+V_a)_{ij}\Phi_j\delta V_a.
\end{equation}
The variation of the trace:

\begin{eqnarray}
\delta\text{Tr}\Psi\Sigma&=&-\frac{1}{2}\bar{D}_+\Psi\delta\left(e^{-V}D_-e^V
\right),\notag\\
&=&\left(2i\epsilon_{abc}\bar{D}_+\Psi_cD_-V_b-i\epsilon_{abc}D_-\bar{D}_+\Psi_cV_b-D_-\bar{D}_+\Psi_a\right)\delta V_a.\\
\delta\text{Tr}\bar{\Psi}\bar{\Sigma}&=&-\frac{1}{2}D_+\bar{\Psi}\delta\left(e^{V}\bar{D}_-e^{-V}
\right),\notag\\
&=&\left(2i\epsilon_{abc}D_+\bar{\Psi}_c\bar{D}_-V_b+i\epsilon_{abc}D_+\bar{D}_-\bar{\Psi}_cV_b-D_+\bar{D}_-\bar{\Psi}_a\right)\delta V_a.
\end{eqnarray}

We can get the equation of motion doing the variation $\frac{\delta L}{\delta V}=0$

\begin{align}
\frac{\delta L}{\delta V_a}=&e^{2QV_0}\bar{\Phi}_i(\sigma_a+V_a)_{ij}\Phi_j+2i\epsilon_{abc}\bar{D}_+\Psi_cD_-V_b-i\epsilon_{abc}D_-\bar{D}_+\Psi_cV_b-D_-\bar{D}_+\Psi_a\notag\\
+&2i\epsilon_{abc}D_+\bar{\Psi}_c\bar{D}_-V_b+i\epsilon_{abc}D_+\bar{D}_-\bar{\Psi}_cV_b-D_+\bar{D}_-\bar{\Psi}_a=0.
\end{align}

\subsection{Dualization in the $a=1$ direction}

In this section we only consider one direction in the  vector superfield dualization. In this case the direction of
the $SU(2)$ generator $\sigma_1$. The vector superfield can be expanded as

\begin{eqnarray}
V&=&\text{Ln}\left[\frac{(\Phi_1+\Phi_2)Z_1+iZ_2(\Phi_1-\Phi_2)}{2e^{2Q_0V_0}\Phi_1\Phi_2(\bar{\Phi}_1+\bar{\Phi}_2)}\right]\\
&=&\text{Ln}\left[\frac{\bar{g}(1+g)Z_1+i\bar{g}(1-g)Z_2}{2e^{2Q_0V_0}|\Phi_2|^2(1+\bar{g})}\right]\\
&=&\text{Ln}\left[\frac{\bar{g}(1+g)Z_1+i\bar{g}(1-g)Z_2}{1+\bar{g}}\right]-\text{Ln}(2|\Phi_2|^2)-2QV_0
\end{eqnarray}

With the following definitions:

\begin{equation}
g=\frac{\Phi_2}{\Phi_1}=\frac{-Z_3\pm\sqrt{Z^2_1+Z^2_2+Z^2_3}}{Z_1-iZ_2},
\end{equation}

and:

\begin{equation}
\bar{g}=\frac{\bar{\Phi}_2}{\bar{\Phi}_1}=\frac{-\bar{Z}_3\pm\sqrt{\bar{Z}^2_1+\bar{Z}_2^2+\bar{Z}^2_3}}{\bar{Z}_1+i\bar{Z}_2}.
\end{equation}

Considering the formulae above we can write the dual GLSM Lagrangian as:

\begin{align}
\tilde{L}&=\int d^4\theta\Bigg\{\bigg[\sqrt{Z_1^2+Z_2^2+Z_3^2}-\frac{1}{2}\text{Tr}\left(\bar{D}_+\Psi e^{-V_1}D_-e^{V_1}+D_+\bar{\Psi} e^{V_1}\bar{D}_-e^{-V_1}\right)\bigg]\Bigg|_1\notag\\
&+\bigg[\sqrt{Z_1^2+Z_2^2+Z_3^2}-\frac{1}{2}\text{Tr}\left(\bar{D}_+\Psi e^{-V_1}D_-e^{V_1}+D_+\bar{\Psi} e^{V_1}\bar{D}_-e^{-V_1}\right)\bigg]\Bigg|_2-\frac{1}{2e^2}\bar{\Sigma}_0\Sigma_0\Bigg\}\notag\\
&+\frac{1}{2}\left(-\int d^2\tilde{\theta}t\Sigma_0+c.c\right).
\end{align}

Expanding the exponential product term as:
\begin{eqnarray}
e^{-V}D_- e^V&=&\left(1-V+\frac{V^2}{2}\right)\left(D_- V+\frac{VD_-V+D_-VV}{2}\right)\notag\\
&=&\left(1-V+\frac{V^2}{2}\right)\left(D_- V+VD_-V\right)\notag\\
&=&D_-V.
\end{eqnarray}

We have that a part of $\tilde{L}$ can be written as

\begin{align}
-&\frac{1}{2}\int d^4\theta \text{Tr}(\bar{D}_+\Psi D_-V-D_+\bar{\Psi}\bar{D}_-V)=-\int d^4\theta(\bar{D}_+\Psi_1 D_-V_1-D_+\bar{\Psi}_1\bar{D}_-V_1)\notag\\
=&-\left(\int d^4\theta D_-\bar{D}_+\Psi_1V_1+\int d^4\theta D_+\bar{D}_-\bar{\Psi}_1V_1\right)
\end{align}

Substituting the value of $V_1$ we get:

\begin{align}
-\Bigg(\int d^4\theta D_-\bar{D}_+\Psi_1\Bigg\{\text{Ln}\left[\frac{\bar{g}(1+g)Z_1+i\bar{g}(1-g)Z_2}{1+\bar{g}}\right]-\text{Ln}(2|\Phi_2|^2)-2Q_1V_0\Bigg\}+c.c\Bigg).
\end{align}

We see that the second term in the previous integral is zero:

\begin{align}
\int d^4\theta D_-\bar{D}_+\Psi_1\text{Ln}(2|\Phi_2|^2)=\int d^4\theta D_-\bar{D}_+\Psi_1\text{Ln}(2\Phi_2\bar{\Phi}_2)
\end{align}

This is due to the fact that the Logarithm of a chiral superfield is a chiral superfield. So finally

\begin{align}
&\int d^4\theta D_-\bar{D}_+\Psi_1\text{Ln}(2\Phi_2\bar{\Phi}_2)=-\frac{1}{4}\int d\theta^+d\bar{\theta}^-D_-\bar{D}_+[ D_-\bar{D}_+\Psi_1\text{Ln}(2\Phi_2\bar{\Phi})]\\
=&-\frac{1}{4}\int d\theta^+d\bar{\theta}^- D_-\bar{D}_+\Psi_1[D_-\bar{D}_+\text{Ln}(2\Phi_2\bar{\Phi}_2)]=0,
\end{align}

\begin{equation}
\text{Ln}(\Phi_2\bar{\Phi}_2)=\text{Ln}\Phi_2+\text{Ln}\bar{\Phi}_2, \hspace{0.1cm}\text{with}\hspace{0.1cm} D_-\bar{\Phi}_2=\bar{D}_+\Phi_2=0.
\end{equation}

Working with the third term we obtain

\begin{align}
\int& d^4\theta D_-\bar{D}_+\Psi_12Q_1V_0=-\frac{1}{4}2Q_1\int d\theta^+d\bar{\theta}^-D_-\bar{D}_+[D_-\bar{D}_+\Psi_1V_0]=-\frac{Q_1}{2}\int d\theta^+\bar{\theta}^-[D_-\bar{D}_+\Psi_1D_-\bar{D}_+V_0]\notag\\
=&\frac{Q_1}{2}\int d\theta^+\bar{\theta}^-[D_-\bar{D}_+\Psi_1\bar{D}_+D_-V_0]=Q_1\int d\theta^+\bar{\theta}^-[D_-\bar{D}_+\Psi_1\frac{1}{2}\bar{D}_+D_-V_0]\notag\\
&=Q_1\int d\theta^+\bar{\theta}^-D_-\bar{D}_+\Psi_1\Sigma_0
\end{align}

The same for

\begin{align}
\int& d^4\theta D_+\bar{D}_-\bar{\Psi}_12Q_1V_0=-\frac{1}{4}2Q_1\int d\theta^-\bar{\theta}^+D_+\bar{D}_-[D_+\bar{D}_-\bar{\Psi}_1V_0]=-\frac{Q_1}{2}\int d\theta^-\bar{\theta}^+[D_+\bar{D}_-\bar{\Psi}_1D_+\bar{D}_-V_0]\notag\\
=&\frac{Q_1}{2}\int d\theta^-\bar{\theta}^+[D_+\bar{D}_-\bar{\Psi}_1\bar{D}_-D_+V_0]=Q_1\int d\theta^-\bar{\theta}^+D_+\bar{D}_-\bar{\Psi}_1\bar{\Sigma}_0.
\end{align}

The dual Lagrangian expressed as:

\begin{align}
\tilde{L}&=\int d^4\theta\Bigg\{\sqrt{Z_1^2+Z_2^2+Z_3^2}-\left[X_1\text{Ln}\left(\frac{\bar{g}(1+g)Z_1+i\bar{g}(1-g)Z_2}{1+\bar{g}}\right)+c.c\right]\Bigg\}\Bigg|_{12}-\int d^4\theta\frac{1}{2e^2}\bar{\Sigma}_0\Sigma_0\notag\\
&+Q\left(\int d\theta^+\bar{\theta}^-D_-\bar{D}_+\Pi_1\Sigma_0+c.c\right)\Bigg|_{12}+\frac{1}{2}\left(-\int d^2\tilde{\theta}t\Sigma_0+c.c\right)\notag\\
&=\int d^4\theta\Bigg\{\sqrt{Z_1^2+Z_2^2+Z_3^2}-\left[X_1\text{Ln}\left(\frac{\bar{g}(1+g)Z_1+i\bar{g}(1-g)Z_2}{1+\bar{g}}\right)+c.c\right]\Bigg\}\Bigg|_{12}-\int d^4\theta\frac{1}{2e^2}\bar{\Sigma}_0\Sigma_0\notag\\
&+Q_{12}\int d\theta^+\bar{\theta}^-\left(X_1\Bigg|_{12}-\frac{t}{2Q_{12}}\right)\Sigma_0+c.c\notag\\
\end{align}

The last term is the twisted superpotential.\\

The first integral is the kinetic term of twisted chiral superfield in component is given by

\begin{equation}
-\frac{1}{2}\int d^2y\left(-K_{ij}\partial_m\Delta_i\partial^m\bar{\Delta}_j+K_{ij}J_i\bar{J}_j\right)
\end{equation}

$K_{ij}$ is the  Kähler  metric and $\Delta$, $J$ are the scalar component and the auxiliary field of the generic twisted chiral superfield. \\

The second integral is the kinetic term of the gauge field;

\begin{equation}
=\frac{1}{4e^2}\left(\frac{D^2}{2}+\frac{v_{03}^2}{2}-|\partial_k\sigma|^2+i\left[\bar{\lambda}^+(\partial_0-\partial_3)\lambda^++\bar{\lambda}^-(\partial_0+\partial_3)\lambda^-\right]\right),
\end{equation}
with $|\partial_k\sigma|^2=\partial_k\bar{\sigma}\partial_k\sigma\equiv-\partial_0\bar{\sigma}\partial_0\sigma+\partial_3\bar{\sigma}\partial_3\sigma$.\\

The expansion of the last integral is:

\begin{align}
=&-Q_0\bigg[\sqrt{2}\bar{\Delta}(D+iv_{03})-\sqrt{2}\bar{\sigma}(2J-ib_{03}+\frac{1}{2}(\partial_0^2-\partial_3^3)\varphi)-\lambda^+(2i\bar{\rho}^--(\partial_0-\partial_3)w^+)\notag\\
&-\bar{\lambda}^-(2i\mu^+-(\partial_0+\partial_3)\bar{\xi}^-)+c.c\bigg]+\left[\frac{t}{2}\left(D+iv_{03}\right)+c.c\right]
\end{align}

The scalar potential from previous contribution is:

\begin{align}
U=K_{ij}J_i\bar{J}_j-\frac{D_0^2}{8e^2}+\sqrt{2}Q_0(\bar{\Delta}_1+\Delta_1)D_0-2\sqrt{2}Q_0(\bar{\sigma}_0J_1+\sigma_0\bar{J}_1)-\frac{1}{2}(t+\bar{t})D_0
\end{align}

Eliminating the auxiliary field $D_0$ with e.o.m we obtain:

\begin{align}
D_0=4e^2\left(2Q(\Delta_1+\bar{\Delta}_1)-\frac{t+\bar{t}}{2}\right).
\end{align}
This for $U$ we have
\begin{align}
U=K_{ij}J_i\bar{J}_j+e^2\left[2Q(\Delta_1+\bar{\Delta}_1)+\frac{t+\bar{t}}{\sqrt{2}}\right]^2-2\sqrt{2}Q(\bar{\sigma}_0J_1+\sigma_0\bar{J}_1).
\end{align}

Solving also with respect to $J_1$ and $\bar{J}_1$, we noted that $K_{2j}J_2\bar{J}_j=K_{3j}J_3\bar{J}_j=0$, $K_{1j}\bar{J}_j=2\sqrt{2}Q_0\bar{\sigma}_0$ for $J_1$ and $K_{i2}J_i\bar{J}_2=K_{i3}J_i\bar{J}_3=0$, $K_{i1}J_i=2\sqrt{2}Q_0\sigma_0$ for $\bar{J}_1$. There is a branch solution for $\sigma_0=\bar{\sigma}=0$. Putting all above in $U$ we get:

\begin{align}
U=e^2\left[2Q(\Delta_1+\bar{\Delta}_1)-\frac{t+\bar{t}}{\sqrt{2}}\right]^2=e^2\left[4Q\text{Re}(\Delta)-\sqrt{2}\text{Re}(t)\right]^2.
\end{align}

So the susy vacua of dual model is reached in the locus,

\begin{equation}
\text{Re}(\Delta)=\frac{\text{Re}(t)}{2\sqrt{2}Q}=\frac{r}{2\sqrt{2}Q}.
\end{equation}

Considering our case of four chiral superfield with charge (1,1,-1,-1), the potential after eliminating the auxiliary fields is:

\begin{align}
U=e^2\left[2Q_1(\Delta_1^1+\bar{\Delta}_1^1)+2Q_2(\Delta_1^2+\bar{\Delta}_1^2)-\frac{t+\bar{t}}{\sqrt{2}}\right]^2=e^2\left[4x_1^1-4x_1^2-\sqrt{2}r\right]^2.
\end{align}

And the susy vacua space will be:
\begin{equation}
r=2\sqrt{2}(x_1^1-x_1^2),
\end{equation}\\
with $\Delta_1^1+\bar{\Delta}_1^1\equiv2x_1^1$, $\Delta_1^2+\bar{\Delta}_1^2\equiv2x_1^2$ and $t+\bar{t}\equiv2r$.\\

\section{Cancelling the semi-chiral superfield dependence of the dual fields}

In this section we explore if the semi-chiral superfield dependence of the non-Abelian T-duality equations of motion can be eliminated by a gauge transformation on the gauged group.

Let us start with the Lagrangian:

 \begin{align}
L&=\int d^4\theta\Big[\bar{\Phi}_i\left(e^{2Q_1V_0+V_1}\right)_{ij}\Phi_j+\bar{\Phi}_p\left(e^{2Q_2V{_0}+ V_2}\right)_{pq}\Phi_q+\text{Tr}\left(\Psi_1\Sigma_1\right)+\text{Tr}\left(\bar{\Psi}_1\bar{\Sigma}_1\right)\notag\\
&+\text{Tr}\left(\Psi_2\Sigma_2\right)+\text{Tr}\left(\bar{\Psi}_2\bar{\Sigma}_2\right)-\frac{1}{2e^2}\bar{\Sigma}_0\Sigma_0\Big]+\frac{1}{2}\left(-\int d^2\tilde{\theta}t\Sigma_0+c.c\right).
\end{align}

The (anti-)chiral superfields transform under a non-Abelian gauge symmetry by:
   
   \begin{align}
   \Phi'=e^{iq\Lambda}\Phi\implies\Phi'^{\dagger}=\Phi'^{\dagger}e^{-iq\Lambda^{\dagger}}.  \label{Phitransfo}
   \end{align}
In the expression the gauge transformation parameter $\Lambda$ is a chiral superfield ($\bar{D}_{\pm}\Lambda=D_{\pm}\Lambda^{\dagger}=0$).
In terms of the gauge gauge generators $T_a$ one can write  $\Lambda_{ij}=\Lambda_aT_{ij}^{a}$. 
For $SU(N)$ the generators $T^{a}$ satisfy a certain Lie algebra.
   
   \begin{equation}
   [T^{a},T^{b}]=if^{abc}T^{c},
   \end{equation}
$f^{abc}$ denotes the group structure constants.  The generators are normalized as Tr $T^{a}T^{b}=k\delta^{ab}$, $k>0$.
   The vector superfield $V$ transforms under the gauge transformations as:
   
   \begin{equation}
   {e^{qV}}'=e^{i\bar\Lambda}e^{qV}e^{-i\Lambda}. \label{Vtransfo}
   \end{equation}
The kinetic term on the Lagrangian is invariant under (\ref{Phitransfo}) and (\ref{Vtransfo}):
   
   \begin{equation}
   \Phi'^{\dagger}{e^{qV}}'\Phi'=\Phi^{\dagger}{e^{qV}}\Phi.
   \end{equation}
   
The twisted field strength $\Sigma$ (fuerza de campo)  and its adjoint are given by:
   \begin{align}
   \Sigma=\frac{1}{2}\bar{D}_+(e^{-V}D_-e^V), \hspace{0.1cm} \Sigma^{\dagger}=\frac{1}{2}D_+(e^V\bar{D}_-e^{-V}).
   \end{align}
Such that under non-Abelian gauge transformations
   
   \begin{align}
   \Sigma'=&\frac{1}{2}\bar{D}_+\big(e^{i\Lambda}e^{-V}e^{-i\Lambda^{\dagger}}D_-(e^{i\Lambda^{\dagger}}e^{V}e^{-i\Lambda})\big)=\frac{1}{2}\bar{D}_+\big(e^{i\Lambda}e^{-V}D_-(e^{V}e^{-i\Lambda})\big),\notag\\
   =&\frac{1}{2}\big(\bar{D}_+(e^{i\Lambda}e^{-V}D_-e^{V}e^{-i\Lambda})+\cancel{\bar{D}_+(e^{i\Lambda}D_-e^{-i\Lambda}})\big),\notag\\
   =&e^{i\Lambda}\Sigma e^{-i\Lambda}.\\
   \implies&\Sigma'^{\dagger}=e^{i\Lambda^{\dagger}}\Sigma^{\dagger} e^{-i\Lambda^{\dagger}}.
   \end{align}
   
One has the following transformation $\Psi'=e^{i\Lambda}\Psi e^{-i\Lambda}$, $\Psi'^{\dagger}=e^{i\Lambda^{\dagger}}\Psi^{\dagger} e^{-i\Lambda^{\dagger}}$  such that:   
   \begin{equation}
   \text{Tr}(\Psi'\Sigma')=\text{Tr}(e^{i\Lambda}\Psi e^{-i\Lambda}e^{i\Lambda}\Sigma e^{-i\Lambda})= \text{Tr}(\Psi\Sigma).
   \end{equation}
   The equations of motion obtained to vary (1) with respect to $V_1$ are:
 
 \begin{align}
\bar{\Phi}_i\left(e^{2Q_1V_0+V_1}\right)_{ij}\Phi_j-\frac{1}{2}\text{Tr}\Big[\Big(\chi_1\tau_1+\tau_1\chi_1+D_-\tau_1\Big)\Delta V_1\Big]-\frac{1}{2}\text{Tr}\Big[\left(\bar{\chi}_1\bar{\tau}_1+\bar{\tau}_1\bar{\chi}_1-\bar{D}_-\bar{\tau}_1 \right)\Delta V^{\dagger}_1\Big]=0,
   \end{align}
with $\chi=e^{-V}D_-e^V$, $\chi^{\dagger}={\color{blue}-}e^V\bar{D}_-e^{-V}$, $\tau=\bar{D}_+\Psi$ y $\tau^{\dagger}=D_+\Psi^{\dagger}$.\\
   
We want a symmetry transformation that allows to eliminate the term $\text{Tr}\big((\chi\tau+\tau\chi)\Delta V\big)$ de las e.o.m.
   
   \begin{align}
   \chi=&e^{i\Lambda}e^{-V}e^{-i\Lambda^{\dagger}}D_-(e^{i\Lambda^{\dagger}}e^{V}e^{-i\Lambda})=e^{i\Lambda}e^{-V}D_-e^{V}e^{-i\Lambda}+e^{i\Lambda}D_-e^{-i\Lambda},\notag\\
   =&e^{i\Lambda}\chi e^{-i\Lambda}+e^{i\Lambda}D_-e^{-i\Lambda}.
   \end{align}
   
   \begin{align}
   \text{Tr}\big((\chi'\tau'+\tau'\chi')\Delta V\big)=& \text{Tr}\big[\big(e^{i\Lambda}\chi \cancel{e^{-i\Lambda}e^{i\Lambda}}\bar{D}_+\Psi e^{-i\Lambda}+e^{i\Lambda}D_-e^{-i\Lambda}e^{i\Lambda}\bar{D}_+\Psi e^{-i\Lambda}+\notag\\
   +&e^{i\Lambda}\bar{D}_+\Psi \cancel{e^{-i\Lambda}e^{i\Lambda}}\chi e^{-i\Lambda}+e^{i\Lambda}\bar{D}_+\Psi \cancel{e^{-i\Lambda}e^{i\Lambda}}D_-e^{-i\Lambda}\big)\Delta V\big],\\
   =&\text{Tr}\big[\big(e^{i\Lambda}\chi\bar{D}_+\Psi e^{-i\Lambda}+e^{i\Lambda}D_-e^{-i\Lambda}e^{i\Lambda}\bar{D}_+\Psi e^{-i\Lambda}+e^{i\Lambda}\bar{D}_+\Psi\chi e^{-i\Lambda}+\\
   +&e^{i\Lambda}\bar{D}_+\Psi D_-e^{-i\Lambda}\big)\Delta V\big],\\
   =&\text{Tr}\big[\big((e^{i\Lambda}\chi-D_-e^{i\Lambda})\bar{D}_+\Psi e^{-i\Lambda}+e^{i\Lambda}\bar{D}_+\Psi(\chi e^{-i\Lambda}+D_-e^{-i\Lambda})\big)\Delta V\big]
      \end{align}
      
      Con $e^{i\Lambda}D_-e^{-i\Lambda}=D_-(e^{i\Lambda}e^{-i\Lambda})-D_-e^{i\Lambda}e^{-i\Lambda}=-D_-e^{i\Lambda}e^{-i\Lambda}$.
      If we want the trace to annihilate we need to solve the equations:
      \begin{eqnarray}
      e^{i\Lambda}\chi-D_-e^{i\Lambda}=0,\label{C16}\\
      \chi e^{-i\Lambda}+D_-e^{-i\Lambda}=0.\label{C17}
      \end{eqnarray}
     But both equations are equivalent. From (\ref{C17}) we observe that:
      \begin{eqnarray}
        \chi e^{-i\Lambda}+D_-e^{-i\Lambda}&=&e^{i\Lambda}\chi e^{-i\Lambda}+e^{i\Lambda}D_-e^{-i\Lambda}=e^{i\Lambda}\chi e^{-i\Lambda}+\cancel{D_-(e^{i\Lambda}e^{-i\Lambda})}-D_-e^{i\Lambda}e^{-i\Lambda},\notag\\
        &=&e^{i\Lambda}\chi-D_-e^{i\Lambda}=0.
      \end{eqnarray}
     We can find a $\Lambda$ satisfying (\ref{C16}) such that the trace annihilates.       
      
      \end{appendices}
\bibliographystyle{utphys}
\bibliography{biblio1}

\end{document}